\documentclass[preprint]{aastex}
\usepackage{epsfig}
\newcommand{\mic}{\hbox{$\mu$m~}} % \mic 
 
\slugcomment{SUBMITTED TO PASP} 
 
\lefthead{Bertero, Boccacci and Robberto} \righthead{Wide-field imaging at mid 
infrared wavelengths:  reconstruction of chopped and nodded data} 
 
\begin{document} 
 
\title{Wide-field imaging at mid-infrared wavelengths  :\\ 
       reconstruction of chopped and nodded data} 
 
\author{M.  Bertero} \affil{INFM and Dipartimento di Informatica e Scienze 
dell'Informazione, Universit\`a di Genova, Via Dodecaneso 35, 
I-16146 Genova, Italy, email: bertero@disi.unige.it} 
\author{P.  Boccacci} \affil{INFM and Dipartimento di Fisica, Universit\`a di 
Genova, Via Dodecaneso 33,\\
 I-16146 Genova, Italy, email: boccacci@fisica.unige.it} 

\and 
 
\author{M.  Robberto\altaffilmark{1}} 
\affil{Space Telescope Science Institute, 
3700 San Martin Dr., Baltimore, MD, 21218, USA, email:robberto@stsci.edu} 
\altaffiltext{1}{On assignment from Astrophysics Division, Space Science Department of ESA}

\begin{abstract}
 
Ground-based astronomical observations at thermal infrared wavelengths  
($\lambda \ge 2.4~\mu$m) 
face the problem of extracting 
the weak astronomical signal from the large and rapidly variable background  
flux. 
The observing strategy most commonly used,  
the so-called ``chopping and nodding'' differential technique, provides  
reliable representations of the target uniquely in the case of compact 
sources while extended and complex sources can be easily   
distorted by their negative counterparts. 
A restoration method, designed to remove the negative values
and to provide reliable representations  of extended  
sources, has been proposed by us in two previous papers 
and validated on simulated images (Bertero et. al 1998, 1999). In this paper
we apply our algorithm to  real images taken at UKIRT telescope 
with the MPIA camera MAX. 
We show that the algorithm successfully 
removes the distortions due to the negative 
counterparts and, in addition, provides noise reduction.
In several cases an enlargement of the field is obtained, in the sense that the
restored larger image provides reliable information on the source structure
outside the central field of view.
The restorations may be affected by artifacts, whose origin can be 
predicted theoretically. We suggest and demonstrate  computational and 
observational procedures for their reduction. 
Once combined with the proper observing strategies,
our inversion method can provide a viable solution to the problem of  
deep imaging of extended sources with large ground-based telescopes. 
 
\end{abstract} 
 
 
\section{Introduction}  
Ground-based astronomical observations at infrared (IR) 
wavelengths can only be performed through a limited number of atmospheric 
windows (\cite{TD93}). Under the best observing conditions, the atmospheric 
transmission closely approaches $\tau \simeq 1$ $(100\%)$ within a few narrow  
spectral intervals, but typical broad-band
values are significantly smaller. Even at excellent sites like Mauna Kea,  
the median transmission in the 8--13~\mic window (N band) is $\tau \simeq 0.85$  
(see e.g. the UKIRT web page).  
 
Besides reducing the number of source photons reaching the 
detector, the atmosphere absorbs and thermally re-radiates isotropically a 
corresponding fraction of energy coming from the space and especially from the 
ground.  In a first approximation, the atmosphere can be conveniently regarded as a 
gray-body radiator with emissivity $\epsilon=1-\tau$ (Kirchoff's law) and 
temperature $\approx 230-250$K.  The corresponding photon flux, peaking at  
10~$\mu$m, is huge compared to the celestial background, e.g. $\approx 10^6$  
times higher than the zodiacal background at 10$\mu$m.  
The telescope itself also adds an important contribution to the background flux.   
Opaque surfaces within the optical beam (secondary mirror spiders, 
primary mirror cell and central obstruction) are near-blackbody radiators,  
whereas the mirrors themselves and the other warm optics, partially reflecting 
or transmitting, are gray-body emitters.   
Proper opto-mechanical design 
and regular cleaning of the optical surfaces  
can reduce the emissivity budget of 
a telescope to a few percents. Still, the total background 
flux at 10\mic within a 1\mic bandpass remains of the order of $10^9$  
photons~s$^{-1}$~m$^{-2}$~arcsec$^{-2}$, roughly corresponding  
to an astronomical source of magnitude $N \approx -3.0$. 
 
Today, it is possible to routinely detect at 10\mic point sources some 12 
magnitudes fainter than the background per square  
arcsecond in a few minutes with 3--4 m  
telescopes and, in particular, UKIRT (\cite{RH98}). To attain 
these performances special observing techniques must be adopted, but  
they are not without drawbacks. 
 
If $x,y$ are angular coordinates in the sky, the signal $s_P$ coming from the 
direction $\{x,y\}$ at time $t$ and detected on the corresponding pixel $P$ of 
the detector can be expressed as:   
\begin{equation}  
s_P=T_P  [ f(x,y) +a(x,y,t)]    
\label{eq:1}  
\end{equation}  
where $f$ is the unknown brightness 
distribution of the celestial source and $a$ is the large and time-variable thermal 
background flux. The transfer function of the detection system  $T_P$ 
includes  the collecting area of the telescope, the field of view of each 
individual pixel and the overall optical transmission. 
Under the conditions described by equation (\ref{eq:1}) 
it is clear that a small error in the estimate of 
$a$ will dramatically affect the extraction of the signal $f$. 
 
The background $a$ can be obtained in principle by pointing the telescope to a 
sky area close to the region of interest at a time $t'$ close to $t$.  Assuming 
that this area corresponds to a shift $\Delta$ in the $y$ coordinate, then the 
new signal $s^\prime_P$ detected at the pixel $P$ is  
\begin{equation} 
s_P^\prime=T_P  [f(x,y+\Delta)+a(x,y+\Delta,t')].   
\label{eq:2} 
\end{equation}  
If the sky area close to that of interest is empty, i.e. 
$f(x,y+\Delta)=0$, and close enough in space and time that the background 
is approximately the 
same, i.e.  $a(x,y+\Delta,t^\prime)\simeq a(x,y,t)$, 
then by subtracting equation 
(\ref{eq:2}) from equation (\ref{eq:1}) one can obtain $T_P f(x,y)$. 
In this way  
the signal can be known within the accuracy of the system response. 
 
In practice, the telescope must sample the two areas 
fast enough that the temporal fluctuations of $a$ are small.  The actual 
frequency depends on various factors: observing 
wavelength, weather conditions, telescope location etc., but is 
typically faster than a few Hz (\cite{k+91}).  Whereas it is, in general,  
impossible to repeatedly move a telescope at  
these frequencies, a single optical element can be rapidly 
``chopped'' between two slightly different positions, allowing  
the detector to see 
two nearby sky areas.  To minimize pupil motion at the cold stop, 
the 
secondary mirror of the telescope is often undersized (so it becomes the exit pupil
of the tescope) and modulated. This classic  
technique  
is called secondary {\it chopping} and the quantity $\Delta$ is 
the {\it chopping throw} or {\it chopping amplitude}.  
Given a certain amplitude, the maximum chopping frequency is constrained by  
the settling time of the secondary mirror structure, typically of the order of 20-50  
milliseconds. 
 
Even neglecting the efficiency loss due to the time spent observing the sky and 
waiting for the secondary mirror to settle, 
this method presents important limitations. 
First, by moving an optical element of the system, the detectors look through
different parts of the optical system (including high emissivity surfaces like
central obstruction, mirror edges, spiders, etc.). Therefore, the resulting
differential image turns out to be affected by a residual background
variation due to the thermal differences between the two beams. 
In other words,  
chopping is equivalent to rapid switching between two different telescopes,  
one $(A)$ for the source and another ${(B)}$ for the sky. 
We will denote by $\Delta a_{AB}$ the residual difference between the 
corresponding background patterns. 
Second, for optical and mechanical reasons the typical chopping throws 
are usually less than 60 arcseconds, possibly much less for 8-m class 
telescopes.  If the source is extended enough, and/or if the telescope 
is sensitive enough, it can be $f(x,y+\Delta) \neq 0$. 
Therefore, the subtraction of equation (\ref{eq:2}) from  
equation (\ref{eq:1}) gives: 
\begin{equation}  
\Delta s_A = s_P - s_P^\prime = T_P [f(x,y) - 
f(x,y+\Delta) + \Delta a_{AB}]  
\label{eq:3}  
\end{equation}  
where with $\Delta 
s_A$ we indicate that the source has been observed with the ${(A)}$ beam. 
 
To remove the term $\Delta a_{AB}$, the so-called {\it beam-switching} or {\it 
nodding} technique is applied:  the telescope is pointed to a different point on 
the sky, so that the source will be observed with the ${(B)}$ beam 
(this maximizes the signal-to-noise ratio). In our 
notation,  the telescope is shifted by  $-\Delta$ arcseconds in the 
$y$ coordinate.  In this way, at the pixel $P$ the signal $s_P^{\prime\prime}$ 
is obtained and, repeating the entire sequence, the result is:   
\begin{equation} 
\Delta s_B = s_P^{\prime\prime} - s_P = T_P  [f(x,y-\Delta) - f(x,y) + 
\Delta a_{AB}].  \label{eq:4}  
\end{equation} 
 
Subtracting equation (\ref{eq:4}) from equation (\ref{eq:3}),  
one gets the so-called {\it chopped and 
nodded image}:   
\begin{equation}  
g_\Delta(x,y) = \Delta s_A - \Delta s_B = T_P  
[- f(x,y-\Delta) + 2f(x,y) - f(x,y+\Delta)],  
\label{eq:5}  
\end{equation}  
i.e. an image which is independent of the atmospheric background and telescope 
thermal pattern.  
If the source brightness distribution is sufficiently compact, 
i.e.  if $f(x,y-\Delta) = f(x,y+\Delta) = 0$, then the problem of extracting 
$f(x,y)$ is solved.  Otherwise a method for recovering $f(x,y)$ 
from the image $g(x,y)$ is required. 
 
As already pointed out (\cite{bec94}, \cite{kae95}), 
there is little doubt that such a  method is becoming highly needed.  
The continuous  
technological developments both in the field of infrared array detectors, 
which allow one to reach the natural background-noise limited performances 
for broad-band imaging, and of large telescope engineering, 
with approximately a dozen of 8-meter class 
telescopes in operation or in an advanced construction phase, are greatly 
improving the sensitivity of thermal infrared instrumentation.  Because 
giant telescopes have smaller pixel scales and have  
limited chopping 
amplitudes, the case $f(x,y-\Delta) = f(x,y+\Delta) = 0$ cannot be regarded 
any more as the typical one so that the $f(x,y \pm \Delta)$  
components of the scene appear as negative signals in the final
image (see equation (\ref{eq:5})). 
In such cases a restoration method is required to obtain, in
the general  case of a structured astronomical object, an image  free from
negative values and reproducing, as far possible, the correct intensity
distribution within the source.
 
In (\cite{rob98}, hereafter BBR98) a preliminary investigation of  
this problem has been performed and it has been found  
that an iterative regularization method, the so-called projected Landweber 
method, can provide a viable solution. 
The method provides a positive restoration of the chopped and nodded image by
removing completely the negative counterparts of compact and extended sources.
The mathematical properties of the 
problem, as well as the validation of the method by means of simulated images,
have been presented in (\cite{ip99}, hereafter BBDR99). 

In Section 2 of this paper we outline the mathematical model we use for
describing the chopping and nodding technique, we discuss the various kinds 
of instrumental errors which are not included in the mathematical
model and we describe 
the iterative restoration method we propose. 
In addition, we discuss the criteria for stopping the iterations, the 
so-called {\it stopping rules}, which are an important issue for the 
correct use of the method.
Finally, we describe three types of artifacts and show how they are related 
to the mathematical properties of the problem.

In Section 3 we apply the method to real astronomical images taken at the
UKIRT telescope. We show that not only the effects of the negative 
counterparts of the sources are successfully corrected, 
but also the noise is reduced, if the number of iterations
is properly chosen. 
We indicate the stopping rule which is the most convenient at the present stage
of our analysis.
We also show that an enlargement of the field is possible, even if the
restoration outside the field is, in general, less accurate than within.
Moreover our restored images provide several examples of the artifacts 
described at the end of Section 2.

In Section 4 we propose three computational 
and observational procedures for the reduction of the artifacts, and 
we show that by applying these methods it is possible to produce
mid-IR images of  
extended sources with high accuracy and sensitivity. 
Finally, in Section 5 we summarize our results and provide a few practical
hints on 
the instrument alignment at the telescope 
that should permit the algorithm to produce the optimal results.

\section{An outline of the restoration method} 
  
We take, for simplicity, 
$T_P=1$ in equation (\ref{eq:5}).  Then by computing the Fourier transform of  
both sides we get  
\begin{equation}  
\hat g_\Delta(\omega_x,\omega_y)=4sin^2({1\over 2} 
\Delta \omega_y)\hat f(\omega_x,\omega_y)  
\label{eq:6}  
\end{equation}  
where $\omega_x,\omega_y$ are the spatial frequencies associated with the   
variables 
$x,y$ respectively.  As already observed by Beckers (\cite{bec94}), equation 
(\ref{eq:6}) shows that the chopped and nodded image does not contain  
information about $\hat f(\omega_x,\omega_y)$ at the frequencies  
$\omega_{y,k}= 2\pi k/\Delta$ $(k=0,\pm 1,\pm 2,\cdots)$, so that the chopping
and nodding technique looks equivalent to the application of a Fourier grating
to the original object. 
However, the  Fourier transform of the measured data is not zero at these 
frequencies because they are contaminated by noise (\cite{kae95}).
As a consequence, the restoration of  $\hat f(\omega_x,\omega_y)$ 
cannot be obtained by
dividing the Fourier transform of the chopped and nodded image
by the factor $sin^2(\Delta\omega_y/2)$.
 
In general Fourier based methods cannot be used for this 
problem because the functions $f$ and $g$ are not defined on the same domain. 
If the region of interest corresponds to the interval $[0,Y]$ in the  
y-variable, 
then the image $g_\Delta(x,y)$, defined on this interval,  
receives contributions from the values 
of $f(x,y)$ on the broader interval $[-\Delta, Y+\Delta]$. 
A method, able to restore $f(x,y)$ on this interval, will provide
an enlargement of the field.
 
Let us assume that the detector plane is partitioned into $N \times N$ pixels, 
with size $\delta \times \delta$, labeled with an index $j$ corresponding 
to the columns and an index $m$ corresponding to the 
rows of the array. 
Since the chopping is in the $y$-direction, it is parrallel to the columns of
the array. Further we assume that the chopping amplitude $\Delta$ is a multiple of 
the sampling distance $\delta$, i.e. $\Delta=K\delta$ with $K$ integer 
and smaller than $N$. Most of the images we consider are $128 \times 128$
and $K$ is typically (but not necessarily) betwen 30 and 50. 

Let $g_{j,m}$ and $f_{j,m}$ be the samples of $g(x,y)$ and 
$f(x,y)$ respectively. For each $j$ the values $g_{j,m}$, with $m$ running  
from $1$ to $N$, form a vector of length $N$ which will be denoted by  
${\bf g}_j$. It receives contributions from $N+2K$ values $f_{j,m}$,  
with $m$ running from $1$ to $N+2K$, which form a vector of length $N+2K$, 
denoted by ${\bf f}_j$ .  The components of ${\bf f}_j$ with 
$m$ running from $K+1$ up to $K+N$ correspond to the sampling points in the 
region of interest, which will be called the {\it observation region}. 
  
Using these notations, equation (\ref{eq:5}) with $T_P=1$ is replaced by  
the following discrete relationship 
\begin{equation}  
g_{j,m}=-f_{j,m}+2f_{j,m+K}-f_{j,m+2K}  
\label{eq:7} 
\end{equation}  
which, by introducing the matrix $[A]$, defined by  
\begin{equation} 
[A]_{m,n}=-\delta_{m,n}+2\delta_{m+K,n}-\delta_{m+2K,n} \ , \ m=1,2,\cdots,N \ \ 
; \ \ n=1,2,\cdots,N+2K  
\label{eq:9}  
\end{equation}  
and called in the following 
the {\it imaging matrix}, can be written in the synthetic 
form  
\begin{equation}  
{\bf g}_j= [A]{\bf f}_j.   
\label{eq:10}  
\end{equation} 
The image restoration problem consists in estimating ${\bf f}_j$ 
(a vector of length $N+2K$), given ${\bf g}_j$ (a vector of length N). 
Since the imaging matrix does not depend on the index $j$, one has 
to solve the same restoration problem for all columns of the image. 
 
There are various kinds of errors that may arise
using the model described by equation (\ref{eq:10}). 
The first is due to the noise, which mostly consists of read-out and background
Poisson noise. Since 
in the case of large background the latter dominates and
can be approximated by white Gaussian noise, the true image will be
given by 
\begin{equation}  
{\bf g}_j= [A]{\bf f}_j + {\bf w}_j  
\label{eq:16}  
\end{equation} 
where ${\bf w_j}$ is a random vector generated by a white Gaussian process. 
When we approximate equation (\ref{eq:16}) with equation (\ref{eq:10}) we 
commit an error which can be amplified by the restoration method 
and therefore must be properly controlled.

A second kind of error occurs when the chopping amplitude 
$\Delta$ is not an integer multiple of the pixel size $\delta$. In such a case 
the simple expression we use for the matrix $[A]$ must be replaced by a 
more complicated one. However, we performed numerical simulations showing that
this effect is not very important when the ratio $\Delta/\delta$ 
is greater than 10, as it is in most practical situations: 
in practice 
the same results can be obtained by using the exact model or the simpler 
one described above, with a value of $K$  
given by the integer closest to ${\Delta}/{\delta}$.

Finally, a third kind of error can be generated by a misalignment 
of the detector 
array with the chopping and nodding direction. This error is the most  
serious one, because it violates the assumption that image restoration can be 
performed column by column. If this occurs, it must be 
corrected by an appropriate rotation of the image.
   
The imaging matrix $[A]$ is rectangular, with $N$ rows and $N+2K$ columns.
A complete
analysis of the mathematical properties of this matrix  is given
in (BBDR99), where it is shown that a unique solution of the restoration 
problem can be obtained if one looks for a positive solution with minimal 
root mean 
square value. However, since the matrix 
$[A]$ is 
ill-conditioned, this solution can be corrupted by an amplified propagation 
of the data noise, so that regularization methods must be used for
controlling this noise propagation (for an introduction to regularization
methods in image restoration see, for instance, \cite{ber98}).

Taking into account that restored images must be
positive and not corrupted by noise amplification, 
we have implemented a particular version of the so-called
{\it projected Landweber method} (\cite{eic92}), proposing  
the following iterative method (BBR98):

\begin{eqnarray}  
\label{eq:19}  
{\bf f}_j^{(0)} & 
= &0 \\ {\bf f}_j^{(k+1)}&=& P_+\left[{\bf f}_j^{(k)}+\tau\left([A]^Tg-[A]^T[A] {\bf 
f}_j^{(k)}\right)\right] \nonumber  
\end{eqnarray}  
where:   
\begin{itemize} 
\item $[A]^T$ denotes the transposed of the matrix $[A]$;  
\item 
$P_+$ is the convex projection onto the closed and convex set of the 
non-negative vectors, defined by 
\begin{eqnarray} \label{eq:20} (P_+{\bf f})_{n} &= & f_n \ \ ,\ \ {\rm if} 
\ \ f_n>0 \\ &= & 0 \ \ ,\ \ {\rm if} \ \ f_n \leq 0 ; \nonumber  
\end{eqnarray} 
\item $\tau$ is a positive parameter, known as {\it relaxation parameter}, 
which, for our problem, must be smaller than $0.125$ (in our code we take 
$\tau=0.1125$).  
\end{itemize} 
 
The implementation of the algorithm defined by equation (\ref{eq:19}) is  
discussed in (BBDR99). 
Here we focus on the fact that  a basic property of the method is that 
it has a regularizing 
effect, usually called {\it semiconvergence}:  the iterates  
${\bf f}_j^{(k)}$ first approach the unknown brightness distribution and then  
move away, because the iterations amplify noise propagation (\cite{ber98}).
For this reason, it is very important to have a criterion for 
selecting the optimal number of iterations in order to obtain 
the best possible approximation of the unknown brightness distribution. 
Below we discuss the criteria which can be used 
to stop the iterations.

We propose two stopping rules based on the 
so-called {\it discrepancy principle} (\cite{ber98}). 
The first works column by column. 
We define the {\it discrepancy} $\varepsilon_j^{(k)}$ between 
the j-th column of the measured data and the j-th column of the data computed 
by means of the k-th iterate as 
the root means square (r.m.s.) value of the vector 
$[A] {\bf f} ^{(k)}_j - {\bf g}_j$:
\begin{equation}
\varepsilon_j^{(k)}=\|[A]{\bf f}_j^{(k)}-{\bf g}_j\|=\left({1\over N} 
\sum_{n=1}^N \mid ([A]{\bf f}_j^{(k)})_n-g_{j,n}\mid^2\right)^{1/2}.
\label{eq:21b}
\end{equation}

From general results  
on the projected Landweber method (\cite{eic92}), it is know  
that $\varepsilon_j^{(k)}$ is a decreasing function of $k$, tending to 
zero when $k\rightarrow \infty$. Then, according to the discrepancy principle,
the iterations can be stopped when  
$\varepsilon_j^{(k)}$ becomes smaller than some estimate $\varepsilon_j$
of the r.m.s. error $\|\bf{w}_j\|$.
In the case of white noise with variance $\sigma^2$, a quite natural estimate
is $\|\bf{w}_j\|\simeq \sigma$, so that the iterations can be stopped when
$\varepsilon_j^{(k)}\leq \sigma$.  
This criterion 
essentially means that one does not look for a very accurate fit of the data 
because, in such a case, one would be looking for a solution fitting not only 
the signal but also the noise. 

Even if the same value of $\varepsilon_j$ is used for all columns, the number
of iterations in general is changing from column to column: the number of
iterations is small if the column is characterized 
by a low value of the signal-to-noise ratio S/N and is larger if S/N is
higher. 
If one does not expect the ratio S/N to change 
dramatically from column to column it may be more convenient to use a second 
stopping rule which provides the same number of iterations 
for all columns. To this purpose we  define the {\it average relative 
discrepancy} as follows
\begin{equation} 
\varepsilon^{(k)}=\left({\sum_{j=1}^{N^\prime} \|[A]{\bf f}_j^{(k)}-{\bf 
g}_j\|^2}\over {\sum_{j=1}^{N^\prime} \|{\bf g}_j\|^2} \right)^{1/2} 
\label{eq:22}  
\end{equation}  
where $N^\prime$ is the number of columns to be restored. The iterations
can be stopped when $\varepsilon^{(k)}$ is smaller than some estimated value
$\varepsilon$ of the relative r.m.s. error affecting the image.
Typically we use a value of $\varepsilon$ corresponding
to a few percent error. If an estimate of $\sigma$ is available, then an 
estimate of $\varepsilon$ can be obtained by replacing each term in 
the numerator of equation (\ref{eq:22}) with $\|{\bf w}_j\|^2\simeq \sigma^2$.
Therefore, if we have two images of the same object with different noise
levels, we must use a larger value of $\varepsilon$, hence a smaller number 
of iterations, for the noisier one.

However, in the case of very noisy images it is important to observe that,
if the value of $\varepsilon$ defined above is used, then the criterion stops
the iterations too early. This effect is due to a property of the discrepancy
principle which has been discovered empirically and is well documented
in the literature (\cite{ber98}). The criterion can be corrected by using
a value of $\varepsilon$ smaller than that defined above 
(for instance by a factor of 2).

In the next Section, by applying both stopping criteria to real images, we 
show that the criterion based on equation (\ref{eq:22}) works
better than that based on equation (\ref{eq:21b}).

As shown in (BBR98) and (BBDR99), using both synthetic and real images, the 
method outlined above provides satisfactory restorations under many  
circumstances. However the restored images may exhibit a few 
typical artifacts. As discussed in (BBDR99)
these artifacts are due to the particular structure of the imaging 
matrix $[A]$ and not to the various kinds 
of error discussed above (noise, non integer chopping, 
misalignment): these artifacts are present also 
in the restoration of synthetic images which are free of these errors 
(see BBDR99). 
For sake of simplicity they can be classified as {\it Type A)}, 
{\it Type B)} and {\it Type C)} artifacts.

{\it Type A) artifacts} - Multiple ``ghost'' images, spaced by $K$, 
of bright stars;  
they may appear as dark images over a bright background or as bright images  
over a dark background. These multiple images are a residual effect of 
the missing frequencies discussed at the beginning of 
Section 2 ( see equation (\ref{eq:6})). 
Their presence means that the restoration method 
does not provide a complete interpolation of the missing frequencies, i.e. 
the Fourier components of the unknown object corresponding to these 
frequencies are not completely restored. Since the missing frequencies 
depend on the chopping amplitude $K$, the simplest way for overcoming this 
difficulty is to use other  images with different 
chopping amplitudes. This raises the question of combining two (or more)
images with different chopping amplitudes to avoid the zeros in
(6) at the spatial frequencies $\omega_{y,k} = \pm 2\pi k/\Delta$.
 
{\it Type B) artifacts} - Regions where the restored image is exactly zero.  
They correspond to the negative counterparts of bright extended sources. If   
these regions contain faint positive sources, these sources can be lost.
 
{\it Type C) artifacts} - Discontinuities of the restored brightness 
distribution at the rows corresponding to the following values of the index 
$n$ in the restored image: $n=$ 
$K_1,K,K+K_1,2K,\cdots,K_1+(q+1)K,(q+2)K$ ( $2q + 4$ jumps),
where $q$ is the integer part of the quotient of the division of $N$ by
$K$ and $K_1$ is the remainder ($N=qK+K_1,K_1<K$). This effect is 
related to the finite size of the image and 
is especially evident when bright parts of the object remain outside 
the observation region.

The characteristics of these artifacts will be discussed through
the next Section,
whereas in Section 4 we shall treat methods for their reduction.

%;;;;;;;;;;;;;;;;;;;;;;;;;;;;;;;;;;;;;;;;;;;;;;;;;;;; 

\section{Application of the method to real images}

%;;;;;;;;;;;;;;;;;;;;;;;;;;;;;;;;;;;;;;;;;;;;;;;;; 

\subsection{MAX observations} 
 
In this section we show the results obtained by applying our algorithm  
to a sample of real 
mid-infrared images. The data, taken in different observing  
campaigns and in part still unpublished, have been obtained using MAX 
(\cite{RH98}), the mid-IR imager developed by the Max-Planck-Institut f\"ur 
Astronomie (MPIA) for the United Kingdom Infrared Telescope (UKIRT).  MAX is 
currently 
equipped with a Rockwell International $128\times 128$ SiAs BIB 
array optimized 
for high-background applications.  
The all-reflective optical design provides a 
scale of $0.27$~arcsec/pixel, corresponding to a field of view of 
$35\times35$~arcsec. 
 
Observations with MAX are usually performed under excellent weather 
conditions and take full advantage from the top-ring and the hexapod 
secondary mirror  
mount with tip-tilt adaptive control developed at the MPIA for UKIRT.  
Since the  
fast-guiding system operates on both chopping beams, MAX routinely provides  
images close to the diffraction limit 
($\lambda/D = 0\farcs54$ at $10\mu$m on UKIRT) on hour-long integration times. 
To reach these performances, the chopping throw and nodding amplitude must be  
finely tuned to 
an integer multiple of $0\farcs84$, the guide camera pixel size, corresponding 
to $\simeq 3$ MAX's pixels  (this value has recently changed to  
$0\farcs94$ - N.Rees, private communication).

The data presented here were generally taken using the standard chopping and  
beam switching technique. However, since  the possible use of the 
inversion algorithm was not considered at the time most of the  
observations were done, the alignment of the array with the chopping  
and nodding directions was never perfectly tuned.  
During image post-processing images were rotated to compensate 
for the original misalignments,
so to satisfy the basic assumption of our restoration 
method. 
Due to excellent cosmetic quality of the detector, MAX raw data required 
minimal cleaning. On the other hand, we did not attempt to correct for 
flat field. 
Accurate flat fielding at mid-IR wavelengths
is known to be a critical issue due to the fast and non-uniform variations 
of the background signal. However, we must notice that for various reasons the 
systematic errors that in principle would require flat-field correction in fact
turn out to be less important in our wavelength range. 
Namely:
\begin{enumerate} 
\item non linearity in pixel response is unimportant due to very 
low dynamic range of the images; 
\item the small fields of view allow to build instruments with a fairly 
uniform detector illumination, i.e. vignetting is normally negligible 
(at least, this is the case of MAX);
\item differences in the detector's pixel-to-pixel response are in general 
less than 5\% and  to some extent mitigated by the oversampling; 
\item for faint sources, the flat-field uncertainty in subtracted images 
can be easily dominated by the background noise.
\end{enumerate}
 
Taking  chopping pairs with chopping throws small enough, we can (almost)
simultaneously compare the flux from bright stars in different parts of the 
MAX's field of view. This check, routinely done at the telescope,
provides results that are consistent within $\approx 1\%$, 
i.e. less than the error typically associated to the absolute flux 
calibration in this wavelengths regime.
It is clear, however, that any reduction of systematic effects through 
accurate flat fielding strategy should further improve the results 
obtained by our reconstruction algorithm.

%;;;;;;;;;;;;;;;;;;;;;;;;;;;;;;;;;;;;;;;;;;;;;;;;;;;; 

\subsection{Bright point sources}  
 
We show first the results obtained by applying the algorithm to  
a bright, isolated point source.
Although our restoration method is not really needed in such a case, we will
use it to investigate his effect on noise propagation and photometric 
accuracy as well as the generation of Type A artifacts.

Figure~\ref{BS}a) shows an image of the bright star BS 1370 obtained 
at UKIRT on 1996 August 26-27 through a broad-N band filter  
($\lambda_{eff}=10.16 \mu$m, $\Delta\lambda=5.20\mu$m). The integration  
times were set to 6.1~milliseconds/frame and 12 seconds total, chopping 
at $\approx 2$Hz with $\Delta=10$~arcseconds throw in the N-S direction 
( $K = 36$ ).   
Image post-processing consisted in filtering out some  
electronic noise in 1 out of 16 preamplifier channels and a counterclockwise  
rotation by $\simeq 2.3^\circ$. After post-processing the 
standard deviation of the noise is estimated using 10 columns in the blank region
of the image. The value we obtain is $\sigma \simeq 10.6$ counts/pixel
and this is the value to be used when applying the stopping rule 
based on equation (\ref{eq:21b}). The corresponding value of $\varepsilon$
to be used according to  equation (\ref{eq:22}),
is about $0.033$. Note that the maximum value of the star
intensity is $2.\ 10^4$ counts. Since the restoration method 
reduces this value by a factor of about 2 (see equation (\ref{eq:5})), 
it should also reduce the noise
by a similar factor.
In the following we denote by $\sigma_r^2$ the {\it variance of the
noise in the restored image}. It is estimated on the same columns used for
the estimation of $\sigma^2$.  

We have applied the method to the image of Figure~\ref{BS}a), using the two
stopping rules introduced in Section 2. 
Using the first one, the 
algorithm performs only one iteration when the column does not contain signal;
the maximum number of iterations is $44$ in the case of the column through the
maximum of the star.
After restoration, the peak value becomes $1.04\ 10^4$ and the variance of the
noise is  
$\sigma_r\simeq 1.7$ counts/pixel. 
Therefore noise is reduced by a factor $6$. This very high noise reduction is due
to the strong smoothing effect of the first iterations of the Landweber
method. As a consequence the high frequency components
of the noise are supressed. Regarding
photometric accuracy the integrated flux (estimated through standard 
multi-aperture photometry) equal to $1.03$ the flux of the original image,
i.e.  photometric accuracy is preserved within $3\%$.

Using the second stopping rule, with $\varepsilon =0.033$, the algorithm 
stops after $13$ iterations. The $\sigma_r$ is smaller than $\sigma$ 
by a factor $2.5$ while photometric accuracy is now preserved within $0.4\%$. We
emphasize that the noise reduction we obtain is just what we expect because, as 
already pointed out above, the restoration algorithm reduces the signal by a factor of 2.

This example clearly illustrates that the method assures noise 
reduction and rather good photometric
accuracy. Both effects depend on the number of iterations and, using our 
two stopping rules, it results that the first one provides an 
excessive noise reduction and a poorer photometric accuracy with respect the
second one. 
For these reasons the second stopping rule seems to be preferable.

Concerning artifacts,
the two deep negatives counter-images created by the chopping  
and nodding technique are completely removed. 
Since the linear scale used in Figure~\ref{BS}b) makes difficult to notice the 
presence of any type of problem we
plot in Figure \ref{BScuts}
the profiles of the original and  restored image along the column 
passing through the star maximum. The profile of the original image has been
divided by 2 for comparison and the range of the ordinates 
reduced to $\pm 10\%$ the stellar maximum.
It can be seen that the stellar profile is reproduced with great accuracy and 
that the two large negative counterparts of the original image are replaced by 
two small positive ghosts, similar to spikes. The integrated photometry of
these spikes is about $3.5\%$ the stellar flux of the restored image.
Besides these spikes two other positive ghosts appear at distances 
$\pm 2\Delta$, outside the original target. Their integrated fluxes are
about $10\%$ the stellar flux. We point out again that the positions 
of these ghosts can be exactly predicted since they are located at distances 
which are multiple of the chopping throw.

According to our stopping rules, changing 
the value of $\varepsilon$ will also change the number of
iterations: more precisely decreasing $\varepsilon$ increases $k$.
The comparison of the results obtained with different number of iterations
allows us to provide the following rule of thumb:
an increase of iterations will cause the noise of the restored image to 
increase, as well as the integrated flux of the star. This is first 
underestimated and then overestimated by the algorithm.

%;;;;;;;;;;;;;;;;;;;;;;;;;;;;;;;;;;;;;;; 

\subsection{Faint point sources}  
 
In order to test the method near the 
sensitivity limit, we apply our algorithm to an image 
containing two faint and isolated point 
sources, the system DH/DI Tau (\cite{M+97}).  
The observational parameters and post-processing  
are similar to those presented in Section 4.2, with a total integration  
time $t_{int}=250$~s. 
Again, since the chopping direction was not perfectly aligned
to the array columns, the image has been properly rotated before applying
the inversion algorithm. 

In our analysis we have considered only the central part of the original image
(more precisely the columns from $26$ to $106$) in order to
eliminate the incomplete lateral columns present in the 
rotated image, which could prevent 
an accurate estimate of  $\varepsilon$. 
After rotation $\sigma$ is about $4$ counts/pixel and the
corresponding value of $\varepsilon$ is about $.5$. 
This large value of $\varepsilon$ implies that we have an example of a very
noisy image.

We have used again the two stopping rules of Section 2 with these values
of $\sigma$ and $\varepsilon$.
The number of iterations allowed by the two criteria 
is now very small (in the case of the second one, just one iteration).
The first criterion provides $\sigma_r=0.6$ and underestimates 
the integrated fluxes of DH Tau
(the bright source) and DI Tau by about $20$\%. 
The second criterion provides again $\sigma_r=0.6$ and underestimates the
integrated flux of DH Tau by about $15$\% while correctly reproduces that
of DI Tau.
Therefore both criteria provide a considerable noise reduction but the second
criterion works better even if the integrated flux of the brighter star is
not really satisfactory.

This result is due to the property of the discrepancy principle discussed in
Section 2.
For this reason we reapplied the algorithm by stopping the iterations 
when $\varepsilon_j^{(k)}
\leq \sigma/2=2$, in the case of equation (\ref{eq:21b}), and when 
$\varepsilon^{(k)}\leq \varepsilon/2=2.5$, in the case of equation 
(\ref{eq:22}).
In this case, the first method provides $\sigma_r=1$ and the integrated 
flux of DH Tau is underestimated by $5$\% while that of DI Tau is
overestimated by $5$\%.
The second criterion provides $\sigma_r=1.2$ and an underestimation of about 
$3.5$\% for DH Tau  and of $0.7$\% for DI Tau.
Again the criterion based on the average relative discrepancy turns out
to provide better results, since both integrated fluxes are 
in nice agreement with the flux measured before the inversion.

The original image (after rotation) as well as the best restored image are 
displayed in Figure~\ref{DITau_rot}. In the original image  DH Tau is the 
brighter source, clearly visible with its two negative counterparts,
while DI Tau is the fainter source $\approx 15$ arcseconds to the S-W.  
The derived fluxes, as reported by \cite{M+97}, are 
$F_N=0.137\pm0.005 $Jy (6.26 mag) for DH Tau and 
$F_N=0.030\pm0.005 $Jy (7.90 mag) for DI Tau. 
The restored  image shows no spikes or ghosts so that
the negative counterparts of the brightest star (DH Tau) 
appear properly cancelled.
 
It must be noted that  
the rotation introduces a certain amount of high-frequency noise filtering  
depending on the rotation angle and origin. To investigate the noise decrease 
independently of the particular rotation/filtering applied, we added, somewhat 
arbitrarily, that amount of uniform Gaussian noise needed to preserve the 
average noise level of the original (before rotation) image.
This new image has $\sigma\simeq 7$ counts/pixel and $\varepsilon 
\simeq 0.6$. By applying the method with the stopping based on the average
relative discrepancy (using a reduction of $\varepsilon$ by a factor of 2)
we obtain a result very close to the previous one still with a 
correct noise
reduction and a non-significant marginal variation of photometric accuracy.

From the analysis of the restoration of isolated 
sources, we conclude that the criterion based on the average relative 
discrepancy provides better 
results. Moreover, in the case of very noisy images, 
it is convenient to reduce by a factor $2$ the value of $\varepsilon$
obtained from the $\sigma$ of the noise.
We will adopt these rules in the following sections.

%;;;;;;;;;;;;;;;;;;;;;;;;;;;;;;;;;;;;;;;;;;;;;;; 
 
\subsection{Bright extended sources}  
 
In this subsection we show how the algorithm performs when bright point sources  
coexist with fainter extended structures, i.e. when the field 
is characterized by high dynamic range signal. The central parsec  
of the Galaxy represents an ideal target in this respect, also  
because most of its prominent features lie within the MAX's field  
of view. The image presented here (Figure~\ref{GClabels})  
has been obtained on the night of 11 April 1998 on UKIRT with  
adaptive tip-tilt correction. We used a ``N-narrow'' filter with  
$\lambda_c =  11.6~\mu$m, $\Delta\lambda = 2.5~\mu$m, chopping throw  
$\approx 10"$ in the north-south direction, and 119.6 seconds total  
integration time on source (i.e. on the positive image).  
The airmass was $z=1.55$.

Once again,
before applying the inversion algorithm, we had care of rotating the image by 
0.7 degrees clockwise in order to align the chopping direction to  
the array columns. The image was also expanded to a $256\times256$ 
size, in order to get a chopping throw $\approx 74.9$ pixels, allowing  
us to run the algorithm with $K=75$.
Due to the complex structure of the source, almost entirely filling the field, 
we were unable to get a meaningful estimate of $\sigma$ using a sufficiently
broad black part of the image.
Given the brightness of the source,
we have applied the criterion based on the average 
relative discrepancy assuming
$\varepsilon=0.08$. The algorithm stops after 63 iterations. 
In Figure~\ref{CGric} we present the reconstruction result 
with two different cuts, in order to evidence the full dynamic range of  
the image.  
 
The reconstructed image clearly shows several details, 
both of the ``northern arm'' and of the ``bar'', which  were  
barely visible or completely lost in the negative features of the  
original image. 
Quite remarkably, the algorithm 
correctly finds the IRS~8 source at the very last rows (top) of the  
reconstructed image. Moreover photometry of sources like IRS~7 
turns out to be  
more reliable once the local patchy negative background has been removed.
 
On the other hand, the reconstructed image is not entirely free from Type A)   
artifacts. 
Ghosts of the brightest sources are clearly visible in the lower and upper  
part of the image. Especially IRS~1 produces a bright artificial spot  
$\approx 5"$ south of IRS~9, but virtually every bright IRS source 
generates some 
signature at the position of his negative counterparts, and in some case 
even at the corresponding multiple distances.  
To illustrate more in detail how artifacts look like,  
we plot in Figure~\ref{GCplot} the vertical cut passing through 
IRS~1 on both the original and reconstructed image.  
The original image (thin line) shows the two negative ``valleys''  
at pixels yr. 20 and yr. 94. At the same places, the reconstructed image  
presents two residual peaks, and especially at pixel yr. 20 the ghost 
feature appears prominent.  

Methods for the reduction of these artifacts will be proposed in the next 
Section.

%;;;;;;;;;;;;;;;;;;;;;;;;;;;;;;;;;;;;;;;;;;;; 
 
\subsection{Faint extended sources} 
 
Here we apply the algorithm  
to the case of diffuse emission completely filling the field. 
It is clear that, in this situation, no algorithm can reconstruct 
the absolute value of the original  
image, lost due to the differential observing technique.   
In Figure~\ref{barra}a) 
we present the image of an area within the Orion nebula taken at UKIRT  
on the night of the 9th February 1997 through the standard  
N-band filter.  
The main observing parameters were: integration time 10~ms/frame and 491.52~s  
total on source, chopping throw $\approx 5$~arcseconds in N-S direction,  
airmass $z\simeq 1.23$. 
 
The field is located approximately 2 arcminutes S-E of the Trapezium stars 
and contains two point sources.
That on top is the proplyd HST8=206-446 (\cite{O+93}, \cite{OW96}).   
The diffuse diagonal feature  
crossing the field is the Orion ``bar'', i.e. the ionization front created 
by the Lyman-C photons produced by the Trapezium stars,  
$\theta^1C$ in particular.

An estimate of the noise has been attempted by analysing 
regions of the original image characterized by a rather uniform intensity 
distribution.
We have obtained $\sigma\simeq 4000$ counts/pixel and the corresponding value
of $\varepsilon$ is $0.2$. Therefore we have another example of a rather noisy
image. For this reason we have used the second stopping rule
with $\varepsilon=0.1$. The corresponding number of iterations turns out to
be  $49$.

The result of the reconstruction, performed  
on the original image without any rotation/filtering, 
is presented in Figure~\ref{barra}b).  It can be seen that 
the restored image results nicely free from Type A) artifacts: both 
stellar images appear properly reconstructed with no relevant ghosts.  
The bar structure turns out to be more detailed, but the effect of its 
negative counterpart on the northern side is a flat reconstruction with 
a large  number of zeros. 
This is a typical example of a  
Type B) artifact better evidenced in Figure \ref{HST8plot} where
we plot vertical cuts through the HST8 maximum both in the original image and 
in two restored images obtained with $\varepsilon=0.2$ and $\varepsilon=0.1$
respectively.
The Type B) artifact corresponding to the large negative counterpart of the 
``bar'' looks the same in the two restorations.
Their comparison allows us to conclude that an increase in
the number of iterations enhances both the star peak and the ``bar''. Again 
the noise is reduced by the restoration algorithm.

%;;;;;;;;;;;;;;;;;;;;;;;;;;;;;;;;;;;;;;;;;;;;;;;;;;;;;;;;;;;;;;;;;;;;;;;;;;;; 

\subsection{Bright sources outside the field}

Before concluding this section, we must refer to that kind of artifacts  
that often arise when the image to be  
restored contains the negative counterpart of bright extended 
sources lying outside the field. At the end of Section 3 they were classified 
as Type C) artifacts. They appear as a series of  
horizontal jumps in the reconstructed images, with periodicity that 
depends on $K$ and $K_1$. 
As an example, we show in Figure~\ref{stripes}a) an image of the 
Orion nebula centered on the dark-silhouette proplyd 167-231 (\cite{RBH99}) 
taken with similar parameters of Figure~{\ref{barra}, except for the 10\arcsec 
chopping throw. 
The corresponding restored image, obtained with average relative discrepancy 
$\epsilon=0.02$, is displayed in Figure~\ref{stripes}b) and 
clearly shows the horizontal jump artifacts. In the next Section we will  
describe a simple method for efficiently reduce this kind of artifacts.

%;;;;;;;;;;;;;;;;;;;;;;;;;;;;;;;;;;;;;;;;;;;;;;;;;;;;;;;;;;;;;;;;;;;;;;;;;;;; 

\section{Artifacts reduction}  
 
We propose three methods which can 
be used to eliminate the artifacts illustrated 
in the previous Section.  The first is both observational and computational
and it is a way for reducing Type A) and Type B) artifacts. The
second is only computational since it is based on  manipulations 
performed on a single chopped and nodded image and acts only on Type C)
artifacts.
The last one is again observational and computational and can be
also used for reducing the Type C) artifacts.

%;;;;;;;;;;;;;;;;;;;;;;;;;;;;;;;;;;;;;;;;;;;;;;;;;;;;;;;;;;;;;;;;;;;;;;;;;; 

\subsection{Inversion of multiple images of the same object}

The analysis of the previous Section indicates that the ghosts of a very
bright source, i.e. Type A) artifacts, or the  
areas with zero value, i.e. Type B) artifacts, are very frequent 
and may reduce the quality of the restorations.  
As we observed in Section 2, these artifacts are due to a 
lack of information in the chopped and nodded image (the missing 
frequencies). Images containing different pieces of information  
can then be restored by means of the projected Landweber  
method and recombined. It is clear that no pair of values of $K$  
should have a common divisor.  
In practice, this procedure is similar to that  
suggested by Beckers (\cite{bec94}), but is completely 
different in what concerns data processing. Note that this approach 
reduces all types of artifacts. 
 
The most simple kind of recombination is to take the arithmetic mean of the 
various restored images over the common domain, coincident with the size 
of the restored image with the smallest value of $K$.  
This approach was applied to simulated images, finding  
that it allows to reduce significantly the restoration 
error (\cite{ip99}).

If more than two images are available, we have verified that the median 
usually provides better results. 
If only two images are available,  
it seems more convenient to take the smallest value. In case of images 
containing only compact sources, this recombination removes completely
the Type A) artifacts. In case of images containing both compact and
extended sources, the bright ghosts over dark background are removed while 
the dark ghosts over bright background are preserved. 
 
In order to validate this approach, four other images of the Galactic Center
(besides that one presented in Section 3.4) 
were taken in April 1998.  
The five images have  $\approx 7\arcsec, 10\arcsec, 
13\arcsec, 17\arcsec, 25\arcsec$  chopping throws, 
always in the N-S direction,
corresponding to $K=29,38,51,67,95$. After the inversion, the size of 
the common region is $128\times 186$ pixel.  Figure ~\ref{fig:CG} shows the 
original images with $7\arcsec, 13\arcsec, 25\arcsec$ chopping throw, 
as well as the corresponding restored  images.  
The ghosts of the brightest point source can be easily identified.  
In Figure ~\ref{fig:CGrisultati} we show the combination of the 5 
reconstructed images using the mean (a) and  
the median (b) of the stack. As we anticipated, the median image  
provides the better reconstruction, just marginally affected by Type A)  
artifacts. 
Figure ~\ref{fig:CGrisultati} shows that only the restoration in the 
observation region is entirely reliable, whereas the
restoration outside strongly depends on the
chopping amplitude.

It is certainly possible to get better results by using more refined 
procedures for recombining the various restored images. 
In particular, one can observe that 
the position of the ghosts can be readily foreseen once the reconstructed 
image is available. 
For each image it is possible to build a map of the  
areas  possibly affected by ghosts where other images 
should better be used, and therefore 
a weight function describing the relative reliability of the various images
may be introduced. 
The investigation of this approach is in progress.

%;;;;;;;;;;;;;;;;;;;;;;;;;;;;;;;;;;;;;;;;;;;;;;;;;;;;;;;;;;;;;;;;;;;;;; 

\subsection{Multiple inversions from the same image}  
 
This is a useful method for reducing the Type C) artifacts, i.e. 
jumps in the brightness distribution of the restored images but
it does not modify the Type A) and Type B) artifacts 
whenever they appear.  It is directly suggested  
by the observation that the positions of the discontinuities are related to  
the values of $K_1$ and $K$.  Since $K$ is fixed, the only way for  
changing these positions is to change $K_1$. This can be obtained by  
reducing the value of $N$, i.e.  by removing a certain number of rows from  
the original chopped and nodded image. 
 
Starting with a chopped and nodded image with columns 
of length $N$ and chopping 
amplitude $K$, by applying the inversion algorithm one obtains a 
restored image 
with columns of length $N+2K$ and jumps at the rows indicated at the end 
of Section 3.  
Removing $M$ rows both from the top and from the bottom of the original 
image  
a reduced image with columns of length $N-2M$ is produced.  
Let us assume $K_1 > 2M$, so 
that the value of $q$ does not change while $K_1$ is replaced by $K_1' = 
K_1-2M$. By applying again the inversion method to the  
reduced image,  one obtains 
a new restored image with columns of length $N+2K-2M$ and jumps at the rows 
$K_1', K, K+K_1',\cdots, K_1'+(q+1)K, (q+2)K$.  The pixels of one column 
of this 
image, with $n$ ranging from 1 to $N+2K-2M$, correspond to the pixels of the 
same column of the previously restored image with $n$ ranging from $M+1$ to 
$N+2K-M$, so that the jumps of the second restored image occur at the rows  
$K_1-M, K+M, K+K_1-M,\cdots, K_1+(q+1)K+M, (q+2)K-M$ of the first one. 
In other words, the effect produced by the reduction of the original 
image is a shift of length $M$ towards the top of the restored image,  
for the jumps of odd order and a shift of length $M$ towards the bottom, 
for the jumps of even 
order.  If the positions of the jumps in the two restored images do not 
coincide, by taking the arithmetic mean of the two images over  
the common domain 
one obtains a reduction of the jumps by a factor of 2. The same result holds  
true also when $2M > K_1$, as it can be deduced by observing that, for the 
reduced image, $q$ is replaced by $q'=q-1$ and $K_1$ by $K_1'=K_1+K-2M$. 
 
This procedure can be extended to the case where several 
reduced images are used in addition to the original one.   
If the total number of 
images is $p$ and if the rows are removed in such a way that  
all positions of the 
jumps in the various restored images are different, then, by taking the 
arithmetic mean of these images over the common domain (which coincides
with that of the shortest restored image) we obtain a reduction of 
the jumps by a factor of $p$. The resulting average image will 
produce a much better result than that obtained by a single inversion. 
 
Figure~\ref{nostripes} shows the result obtained by applying this procedure 
to the image presented in Figure~\ref{stripes}a). 
The average image is the arithmetic mean of 
six restored images, one directly obtained from the original image and 
five obtained by removing respectively 1,2,3,4, and 5 rows 
from the top and the bottom of the original one.  The jumps 
are considerably attenuated while the multiple ghosts of the bright 
stars remain unchanged.

%;;;;;;;;;;;;;;;;;;;;;;;;;;;;;;;;;;;;;;;;;;;;;;;;;;;;;;;;;;;;;;;;;;;;;;;;; 

\subsection{Inversion of pasted images}

The Type C) artifacts in the restored image do not appear or, 
at least, are very 
weak if no significant extended source exists outside the observation region.  
However, it is in general possible to take images 
of adjacent regions in order to build a mosaic encompassing the brightest  
sources of the region. If no significant extended source exists 
below and above the mosaic, the algorithm can be run on the combined image 
to obtain a better reconstruction. 
 
We have tested this approach on a couple of images of the giant HII 
region W51. 
The observations were carried out with MAX on August 29-30 1997 through  
the broad N-band filter. The chopping frequency was 2.2~Hz and the chopping  
throw $\approx~30\arcsec$ in the N-S direction. The integration time was set  
to 10.2~ms per frame and 40 seconds total on source (\cite{LRH99}).

The two images (Figure~\ref{w51c}) are centered respectively on W51~IRS1
(Figure~\ref{w51c}b))
and 29~arcseconds north of W51~IRS1 (Figure~\ref{w51c}a)). 
Figure~\ref{w51c}b) shows that, due to 
the large chopping throw, the bright star on top has a negative  
counterpart close to the bottom of the image. 
Another negative counterpart is visible nearby, $\approx 8\arcsec$ to the  
south of  IRS~1. From this  
image alone there is no way to specify if this second source lies above or  
belove the field. Figure~\ref{w51c}a) reveals that it lies above.  
 
The reconstruction of these two images produces the results displayed in  
Figure~\ref{w51cr}a) and Figure~\ref{w51cr}b) respectively. 
These results are clearly unacceptable.  
On the other hand, the two images can be combined to form a $128\times 224$ 
pixel mosaic (Figure~\ref{w51c}c)).   
Figure~\ref{w51cr}c) shows the corresponding reconstruction 
with dynamic range  emphasizing the lowest counts. With respect to  
Figure ~\ref{w51cr}a,b), the improvement in image is striking: the horizontal  
discontinuities within the central field have disappeared, and also the other 
artifacts are significantly reduced. Only discontinuities at the border of
the observation region defined by the mosaic are still visible.   
Note that the ``hole'' present between IRS1 and the north rim 
is real, as near-IR data clearly indicate the presence of a  
dark ridge North of IRS1. 
Note also the source  
$\approx 25\arcsec$ to the south of IRS 1, visible in the 
20~$\mu$m map of \cite{Genzel+82} and correctly  
reconstructed by our algorithm.

%;;;;;;;;;;;;;;;;;;;;;;;;;;;;;;;;;;;;;;;;;;;;;;;;;;;;;;;;;;;;;;;;;;;;;; 

\section{Comments and Conclusions}

We have considered the problem of the reconstruction of astronomical  
data taken at mid-IR wavelengths
in chopping and nodding mode. Studying the mathematical  
properties of the corresponding under-determined linear system, we have  
proposed an iterative method for approximating the positive solution of  
minimal r.m.s. value.
We have implemented the algorithm and tested it on astronomical 
data taken in various observing runs at the UKIRT telescope with the  
MAX camera. We have investigated the nature 
of the artifacts affecting the restored images and suggested various  
computational and observational strategies for their reduction. We find that  
if an extended source is observed with a few different chopping throws,  
possibly building a mosaics if the source extends beyond 
the detector field of  
view, our algorithm can provide a reliable reconstruction of the 
source brightness. 
 
We finally remark that on the basis of the MAX experience the proper  
setup of the camera at the telescope 
is crucial to get the best results from our reconstruction method.  
It can be useful in this respect to provide our check-list for the  
setup of MAX on UKIRT. 

Assuming that the chopping-nodding direction has been chosen to be N-S:
\begin{enumerate} 
\item 
Select chopping throws close to a common multiple of the instrument 
pixel size and of the guide/tip-tilt camera pixel size. 
Use preferably chopping throws 
larger than 1/6 the array size for best reconstruction (see BBDR99). 
\item 
Align the chopping direction to the detector orientation: point to  a bright 
star at the edge  
of the field, and make sure that for large chopping throws 
$x$ coordinates of the positive and negative  
centroids are as close as possible within the  
same column. The chopping direction will be in general different from 
the real N-S direction.  
\item 
Align the nodding direction to the chopping direction. If the telescope 
software 
implements the nodding as a simple jump of the telescope to an 
``offset'' position, be sure that  
both the right ascension and declination (or amplitude and position angle) 
of the offset beam have been given. 
Note that  both values will change with the chopping throw.  
\item 
If the guide camera/cross-head moves when the telescope is pointed to the 
offset beam,  
make sure that also for the camera offset both  right 
ascension and declination have been  introduced.  
\end{enumerate} 
There is a last point that deserves some attention as a potential limit to  
the accuracy of the method: field distortion.  
Off-axis reflectors are a preferred choice for mid-IR imagers, 
as they allow a compact and achromatic optical design. However, these systems  
are in general strongly affected by field distortion. Optical designers often 
trade 
field distortion for the ultimate optical quality. For thermal-IR work, it 
must be kept  
in mind that is still very difficult to obtain a good astrometric calibration,
given also the scarcity of good calibration fields in this  
wavelengths regime. 
 
Our code is free and is available on request to the authors. Two versions  
are available, in C and in IDL. For the IDL version, the typical execution  
time for the inversion of a $128\times128$ image is of the order of a few  
tens of seconds on a Ultra~10 SPARC station.    
%;;;;;;;;;;;;;;;;;;;;;;;;;;;;;;;;;;;;;;;;;;;;;;;;;;;;;;;;;;;;;;;;;;;;;;;;; 

\section{Acknowledgements} 
 
This work was partially funded by the Italian CNAA (Consorzio Nazionale per 
l'Astronomia e l'Astrofisica) under the contract nr. 16/97.

The authors are indebted to S.Ligori for providing the W51 images in advance  
of publication. M.R. acknowledges Steven Beckwith and Tom Herbst for 
discussions on this subject during the long nights spent observing on UKIRT.

%;;;;;;;;;;;;;;;;;;;;;;;;;;;;;;;;;;;;;;;;;;;;;;;;;;;;;;;;;;;;;;;;;;;;;;;;;; 

%;;;;;;;;;;;;;;;;;;;;;;;;;;;;;;;;;;;;;;;;;;;;;;;;;;;;;;;;;;;;;;;;;;;;;;;;;;;;

\newpage

\begin{figure}[ht]  
\begin{center}  
\begin{minipage}{5.cm}  
\begin{tabular}{c} 
\psfig{file=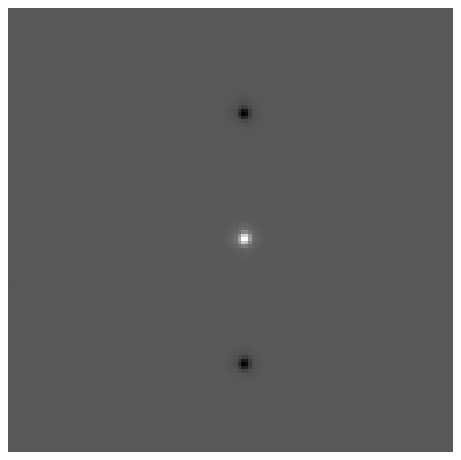,width=5.cm}\\  a) \\  
\end{tabular} 
\end{minipage}  
\begin{minipage}{5.cm}  
\begin{tabular}{c} 
\psfig{file=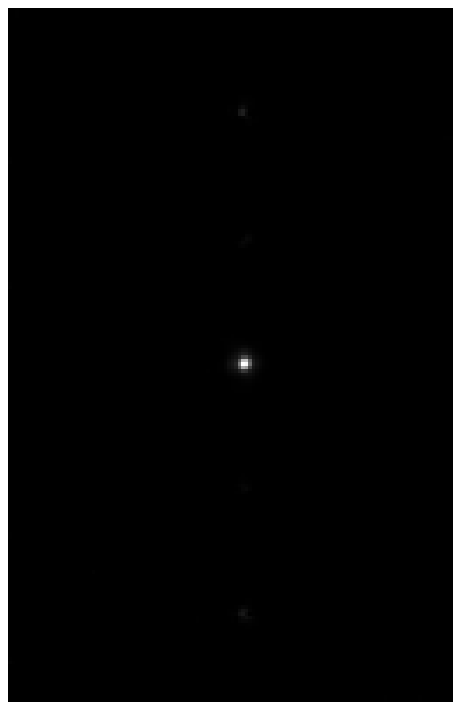,width=5.cm}\\  b) \\  
\end{tabular} 
\end{minipage}  
\end{center}  
\caption{a) Chopped and nodded image of the bright star BS~1370; the black 
circles are the negative counterparts of the star, generated by the chopping 
and nodding technique. b) The restoration of image a) obtained by means 
of our method, using the second stopping rule based on the average relative
discrepancy; the negative counterparts have 
been completely removed (linear-scale representation).}  
\label{BS}  
\end{figure} 

\begin{figure}[ht]
\begin{center}
\psfig{file=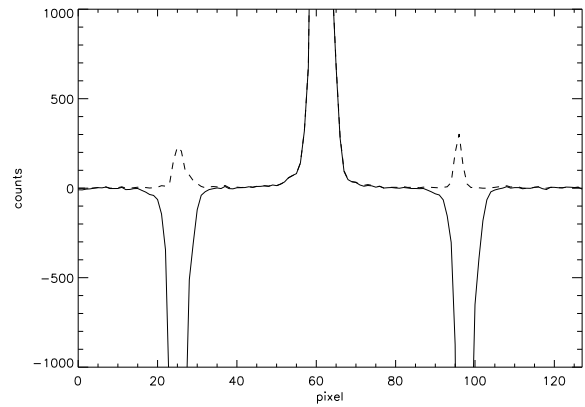,width=8.cm}
\end{center}  
\caption{a) Vertical cuts passing through the maximum of BS~1370 : 
original image (full line), reconstructed image (dotted line).
For comparison, the profile of the original image has been divided by 2. }  
\label{BScuts}  
\end{figure} 

\begin{figure}[ht]  
\begin{center}  
\begin{minipage}{3.25cm}  
\begin{tabular}{c} 
\psfig{file=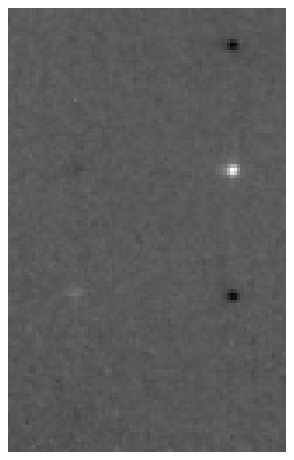,width=3.25cm}\\ a) \\  
\end{tabular} 
\end{minipage}  
\begin{minipage}{3.25cm}  
\begin{tabular}{c} 
\psfig{file=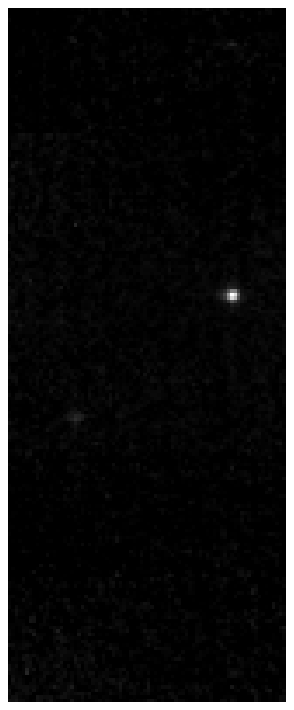,width=3.25cm}\\ b) \\  
\end{tabular} 
\end{minipage}  
\end{center}  
\caption{a) Original image of the system DH/DI Tau (80 columns). 
b) Restored image obtained using the second stopping rule of Section 2 with
$\varepsilon=0.25$ (linear-scale representation).}  
\label{DITau_rot}  
\end{figure} 

\begin{figure}[ht]  
\begin{center}  
\psfig{file=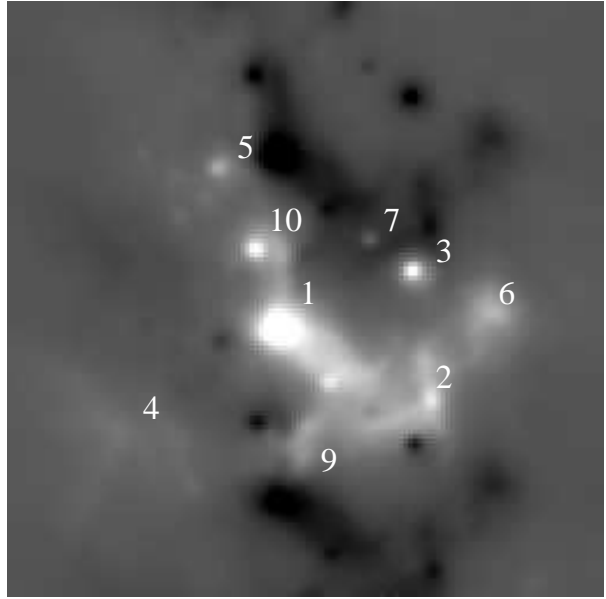,width=8.cm} 
\end{center}  
\caption{Raw N-band image of the Galactic Center with 10\arcsec chopping throw.
The compact sources are labeled according to Becklin et al. (1978). The 
extended black regions are negative counterparts of the extended bright 
sources.} 
\label{GClabels}  
\end{figure} 

\begin{figure}[ht]  
\begin{center}  
\begin{minipage}{5.cm}  
\begin{tabular}{c} 
\psfig{file=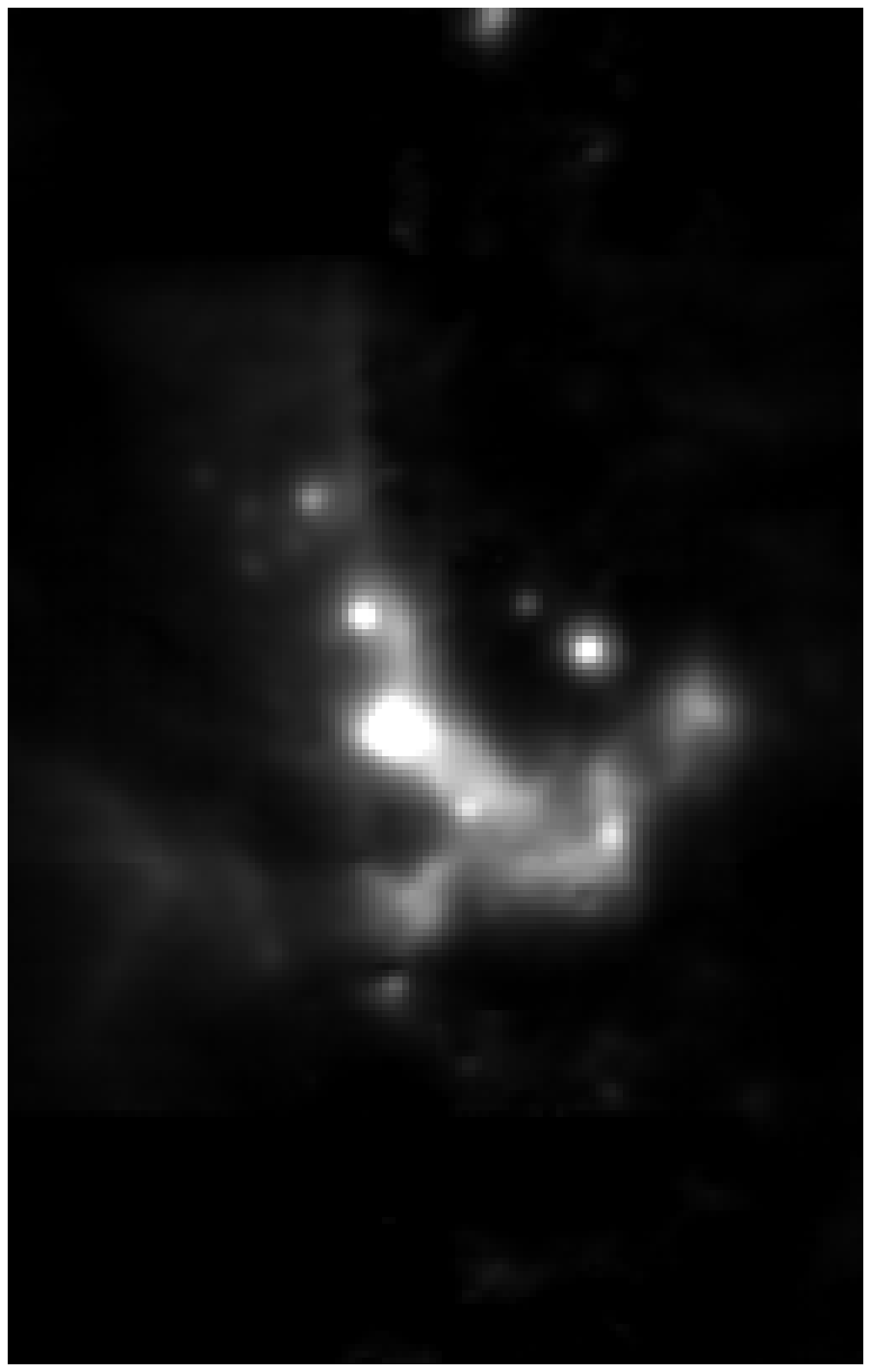,width=5.cm}\\  
\end{tabular} 
\end{minipage}  
\begin{minipage}{5.cm}  
\begin{tabular}{c} 
\psfig{file=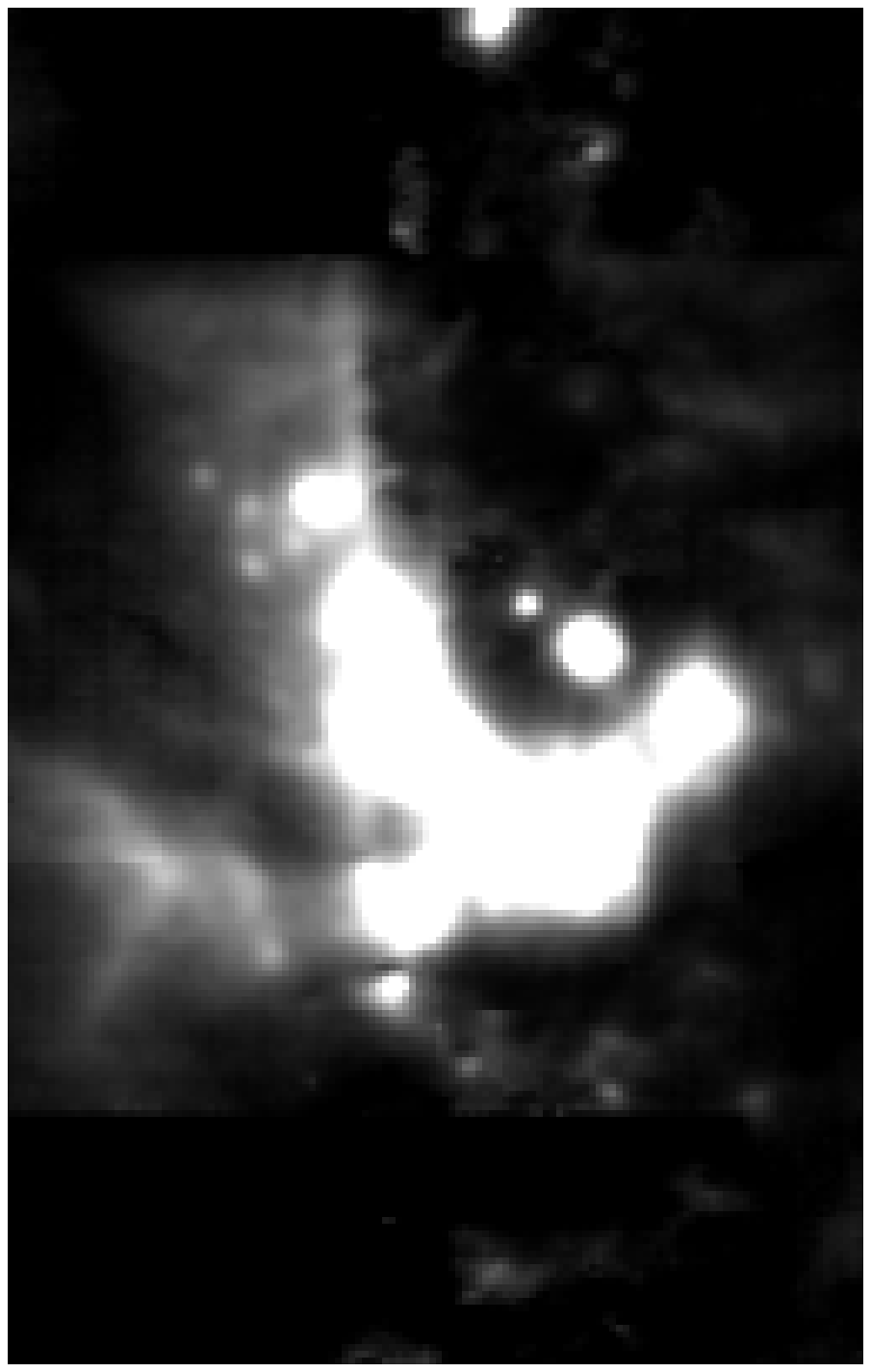,width=5.cm}\\  
\end{tabular} 
\end{minipage}  
\end{center}  
\caption{Restored images of the Galactic Center. The two pictures are  
obtained with two different cuts in the gray scale to evidence the full 
dynamic range of the image.}  
\label{CGric}  
\end{figure} 

\begin{figure}[ht]  
\begin{center}  
\psfig{file=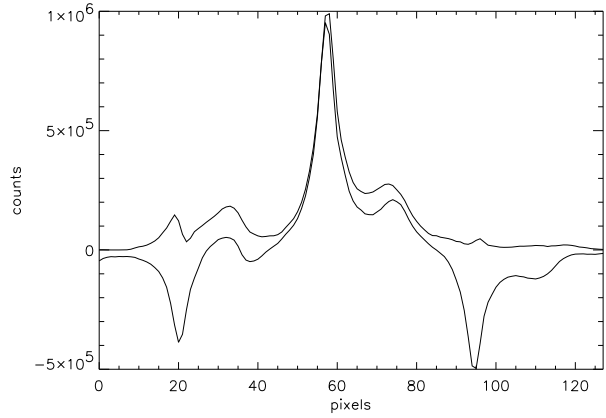,width=8.cm} 
\end{center}  
\caption{Vertical cuts through the IRS 1 source in the original image 
(thin line) and in the reconstructed image (thick line). Pixel 0
is at the bottom of the image.} 
\label{GCplot}  
\end{figure} 

\begin{figure}[ht] 
 \begin{center} 
  \begin{minipage}{5.cm} 
   \begin{tabular}{c} 
\psfig{file=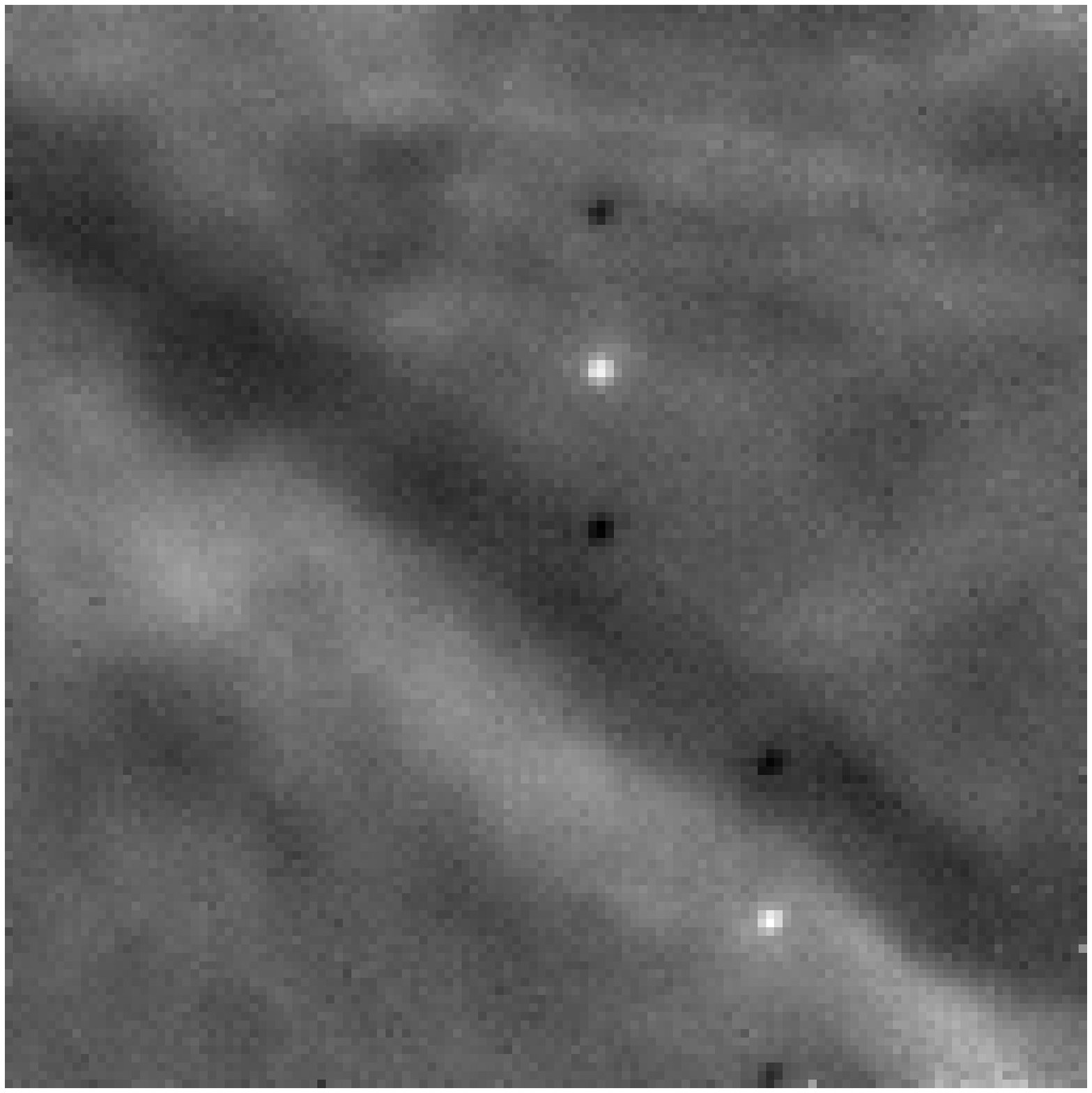,width=5.cm}\\ a)\\ 
   \end{tabular}  
  \end{minipage} 
  \begin{minipage}{5.cm} 
   \begin{tabular}{c} 
\psfig{file=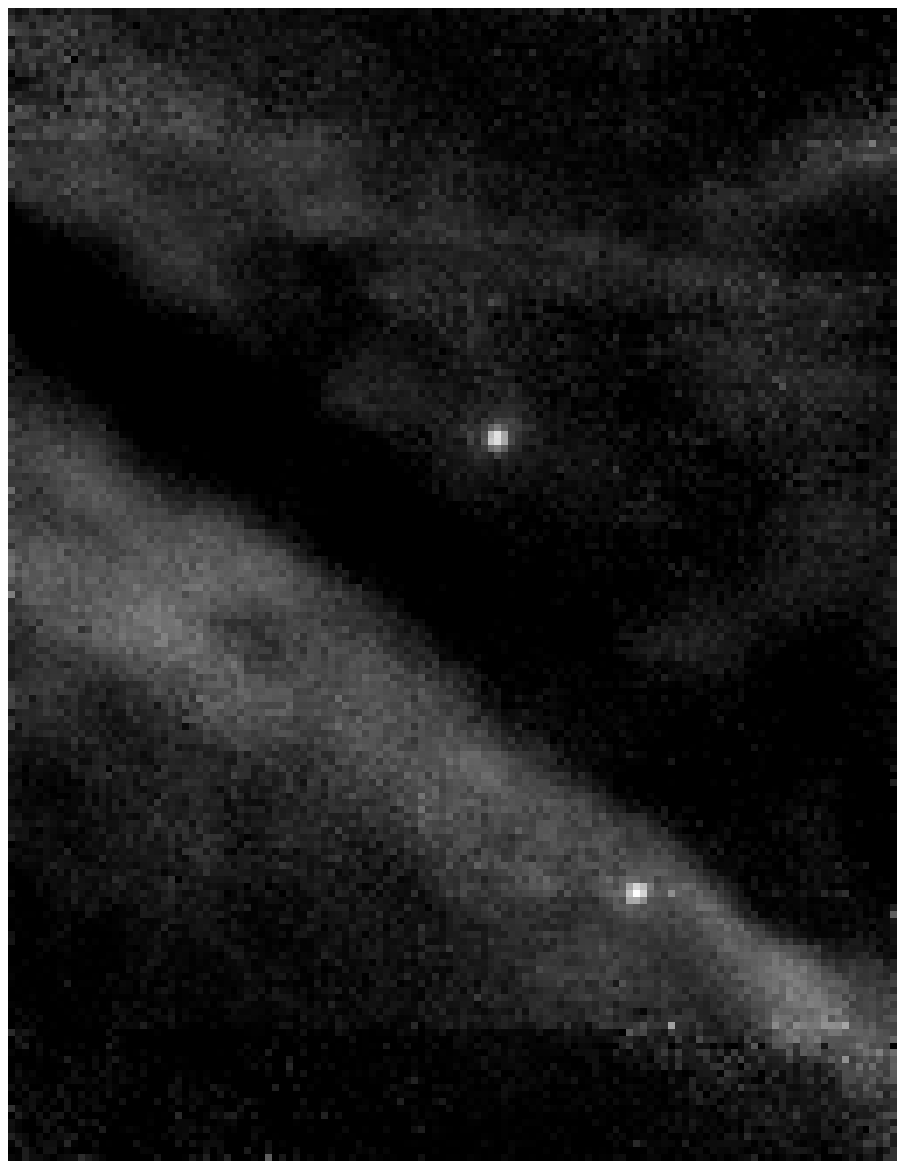,width=5.cm}\\ b)\\ 
   \end{tabular} 
  \end{minipage} 
 \end{center}   
 \caption{a) Raw image of the Orion bar with 5\arcsec chopping throw.
b) The restored image obtained after $49$ iterations ($\varepsilon=0.1$)
(linear-scale representation).} 
 \label{barra}  
\end{figure}

\begin{figure}[ht]  
\begin{center}  
\psfig{file=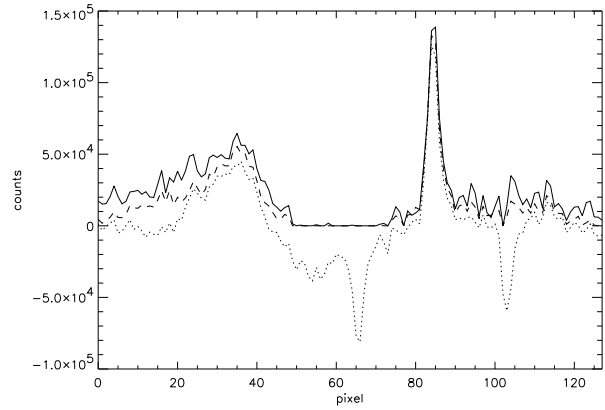,width=8.cm} 
\end{center}  
\caption{Vertical cuts passing through the HST8 maximum for the original image
 (dotted line), the restored image with $\varepsilon=0.2$ (dashed line), 
obtained after $9$ iterations, and the restored image with
$\varepsilon=0.1$ (full line), obtained after $49$ iterations. } 
\label{HST8plot}  
\end{figure} 

\begin{figure}[ht]  
\begin{center}  
\begin{minipage}{5.cm}  
\begin{tabular}{c} 
\psfig{file=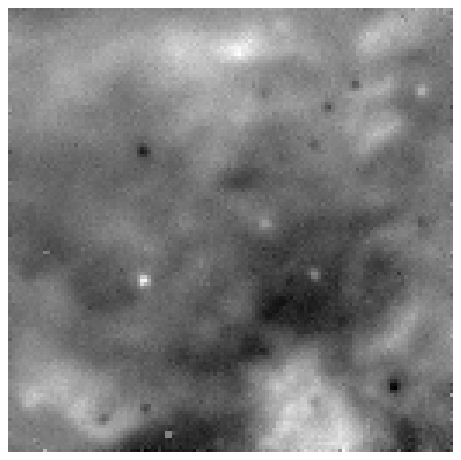,width=5.cm}\\ a)\\  
\end{tabular} 
\end{minipage}  
\begin{minipage}{5.cm}  
\begin{tabular}{c} 
\psfig{file=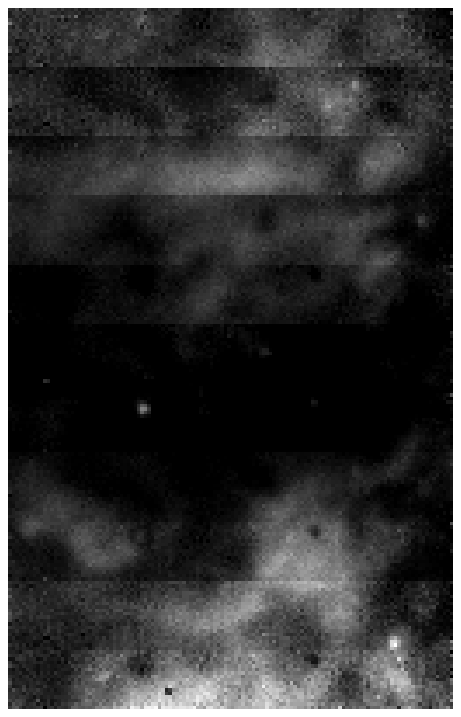,width=5.cm}\\ b)\\  
\end{tabular}  
\end{minipage}  
\end{center}  
\caption{a) Raw image of the Orion nebula, centered on proplyd 167-231, with 
10\arcsec chopping throw. b) The restored image with  
discrepancy $0.02$ (linear-scale representation).}  
\label{stripes} 
\end{figure} 

\begin{figure}
\begin{center} 
\begin{tabular}{c c c }
\psfig{file=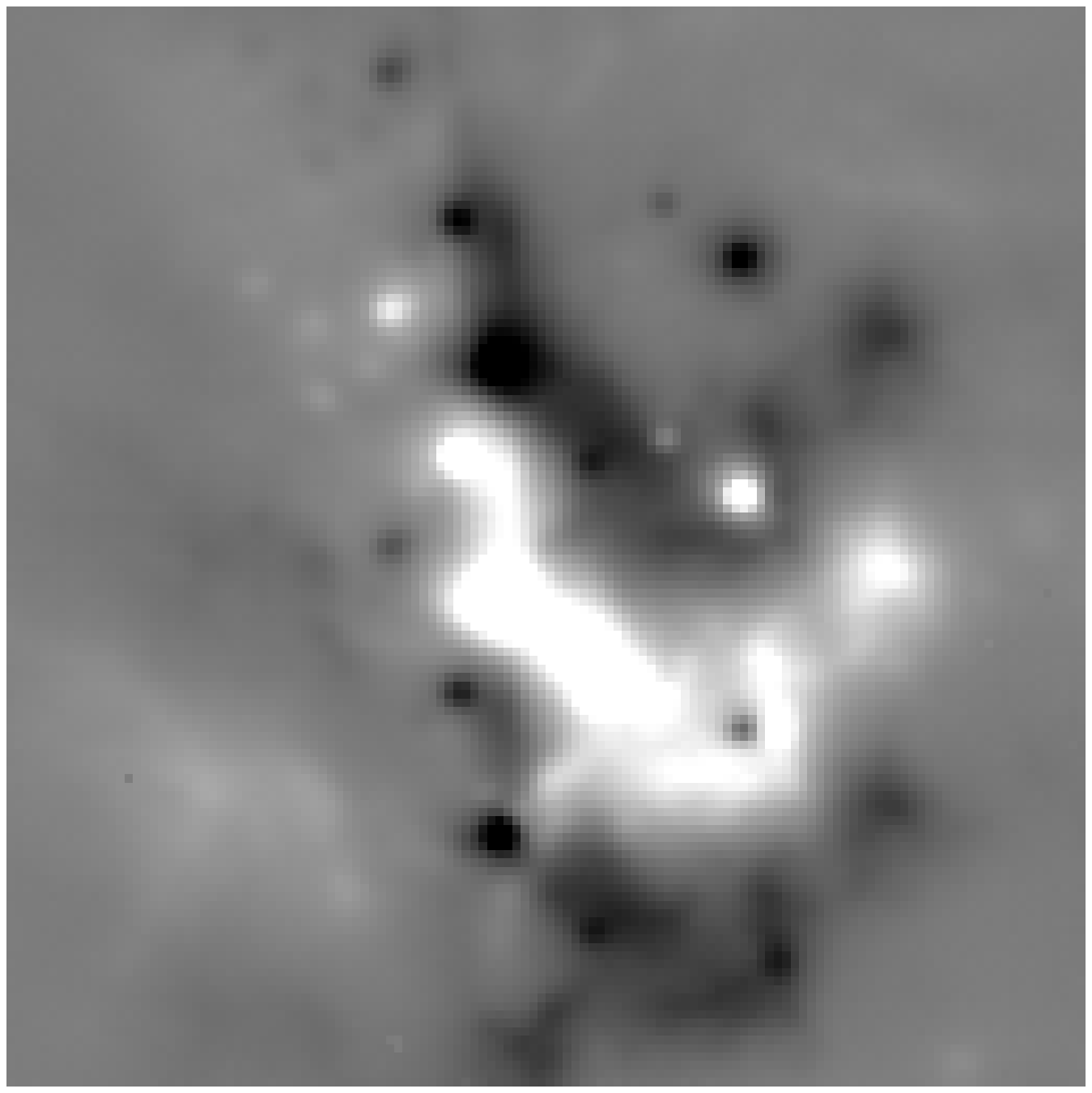,width=3.5cm}&
\psfig{file=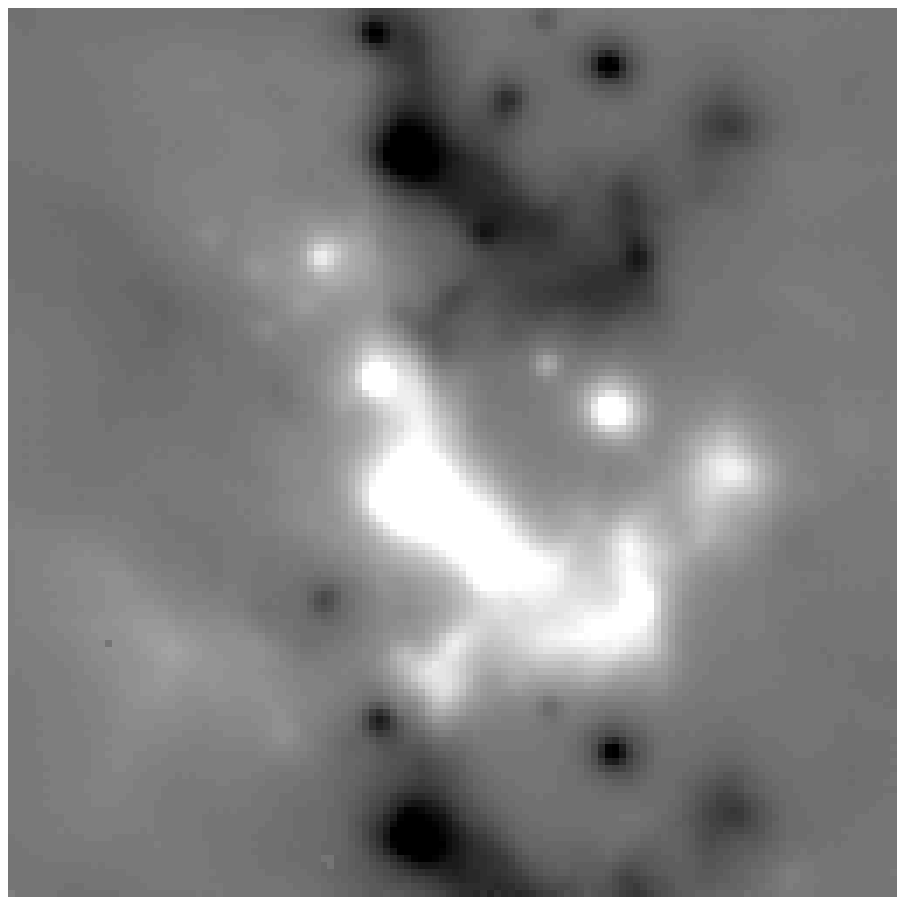,width=3.5cm}&
\psfig{file=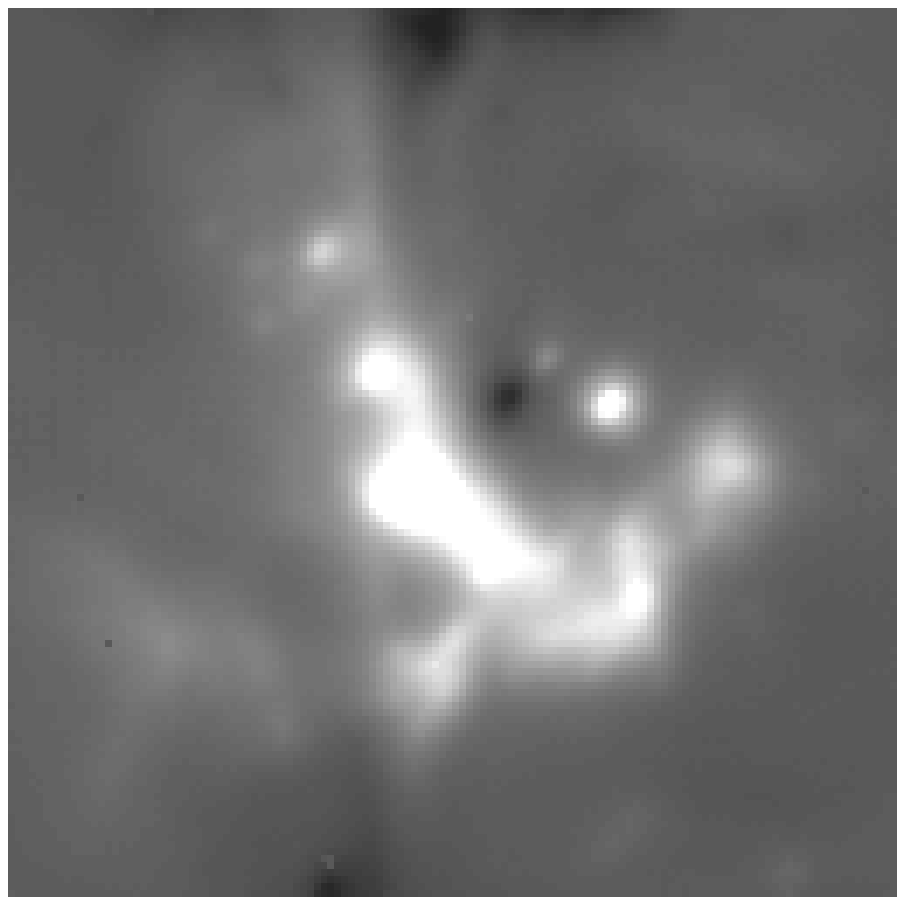,width=3.5cm}\\
a) & b) & c) \\
\end{tabular}
\end{center}
\begin{center}
 \begin{minipage}{3.5cm} 
\begin{tabular}{c}  
\psfig{file=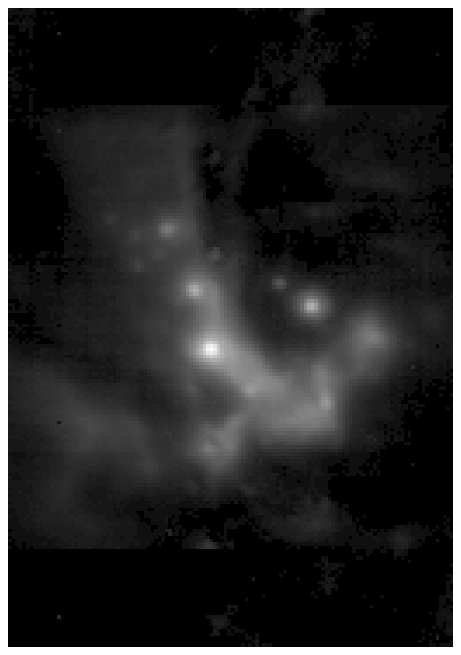,width=3.5cm}\\ d)\\
\end{tabular}
\end{minipage} 
\begin{minipage}{3.5cm} 
\begin{tabular}{c}  
\psfig{file=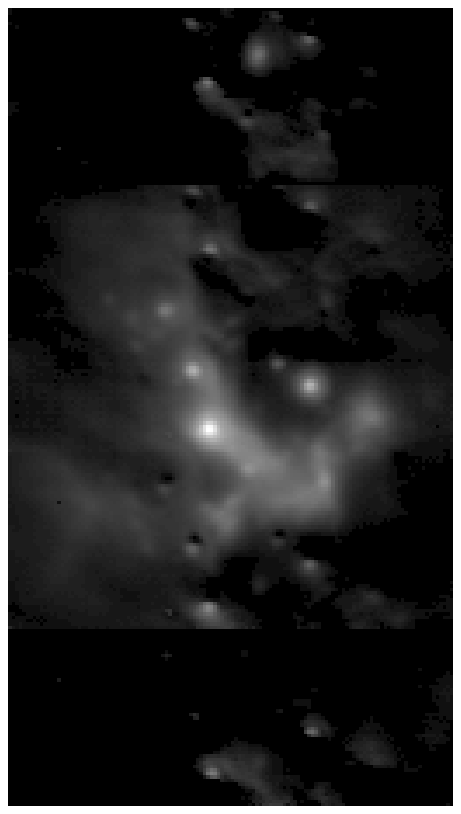,width=3.5cm}\\ e)\\
\end{tabular}
\end{minipage}
 \begin{minipage}{3.5cm} 
\begin{tabular}{c}  
\psfig{file=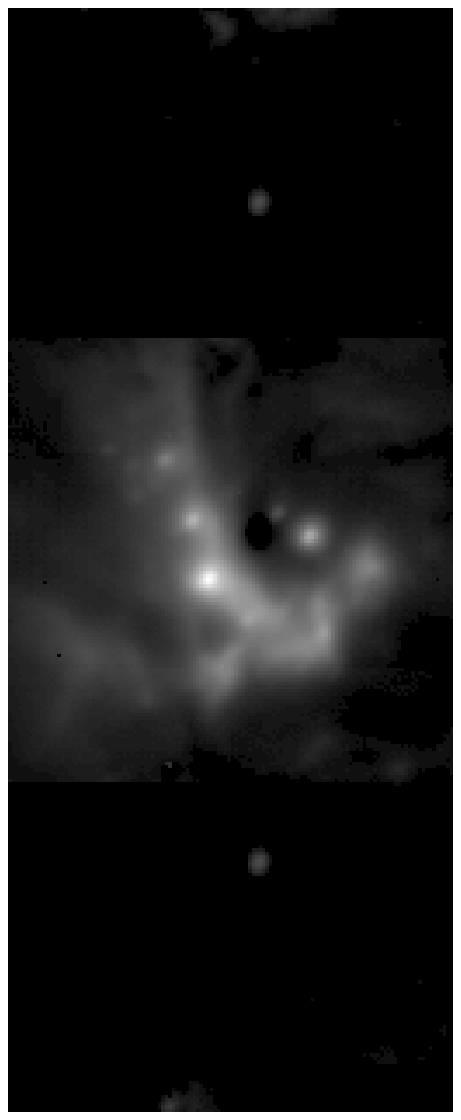,width=3.5cm}\\ f)\\
\end{tabular}
\end{minipage}
\end{center}
\caption{Chopped and nodded images of the galactic center with: 
a) $7\arcsec$ chopping throw, $K=29$; b) $13\arcsec$ chopping throw, $K=51$;  
c) $25\arcsec$ chopping throw, $K=95$. d), e), f)  are the restorations
of a), b) , c) respectively (the restorations are represented using 
square root scale).} 
\label{fig:CG}
\end{figure}

\begin{figure}
\begin{center} 
\begin{tabular}{c c} 
\psfig{file=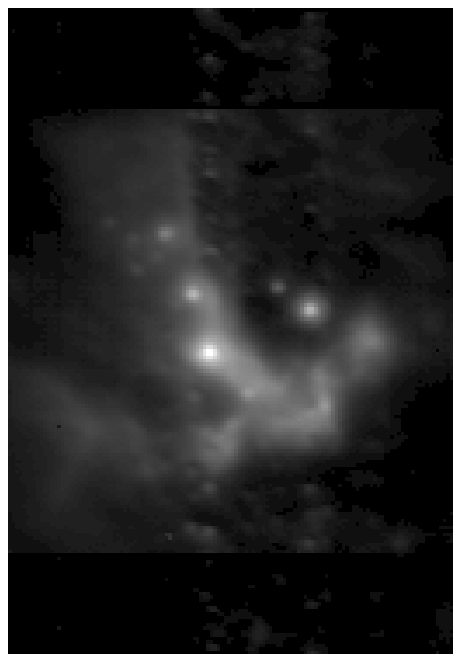,width=4.cm}&
\psfig{file=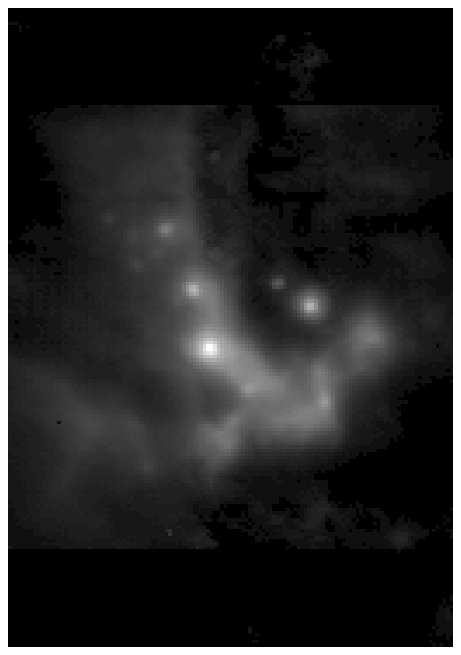,width=4.cm}\\ 
a) & b)\\
\end{tabular} 
\end{center} 
\caption{Recombined images of the Galactic Center, obtained  
taking: a) the arithmetic mean, b) the median of five  
restored images; three of them are shown in Figure~\protect{~\ref{fig:CG}})
(square root scale representation).}
\label{fig:CGrisultati} 
\end{figure}

\begin{figure}[ht] 
\begin{center} 
\psfig{file=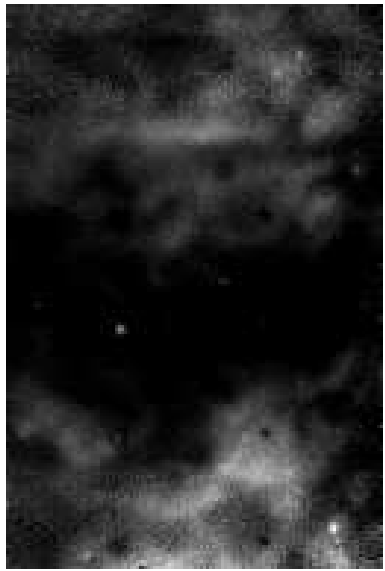,width=5.cm}\\   
\caption{Reconstruction of the image of  Figure \protect{~\ref{stripes}}a) 
obtained by means of the reduction and averaging procedure 
described in the text (linear-scale representation).}  
\label{nostripes} 
\end{center} 
\end{figure} 

\begin{figure}[ht] 
 \begin{center} 
 \begin{minipage}{5.cm} 
   \begin{tabular}{c}   
\psfig{file=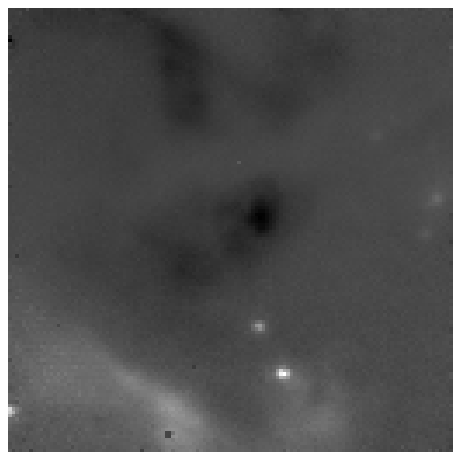,width=5.cm}\\ a) \\ 
\psfig{file=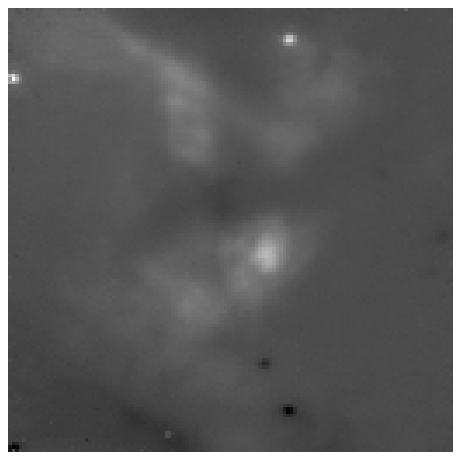,width=5.cm}\\ b) \\  
   \end{tabular} 
  \end{minipage} 
 \begin{minipage}{5.cm} 
   \begin{tabular}{c} 
\psfig{file=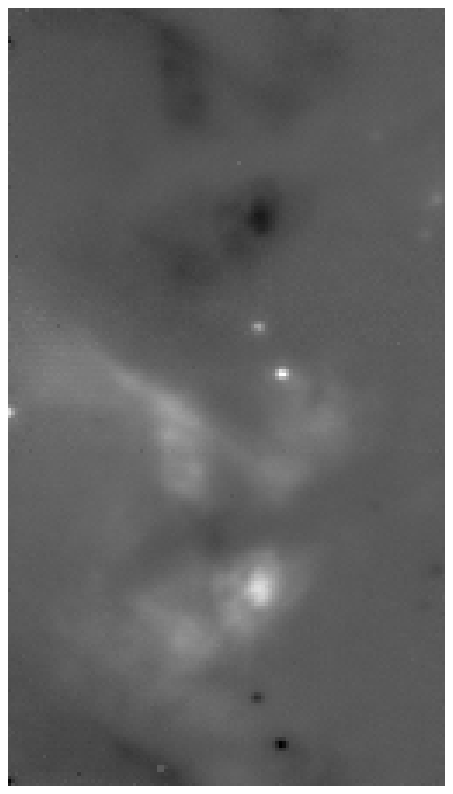,width=5.cm}\\ c) \\ 
   \end{tabular}  
  \end{minipage}
 \end{center} 
\caption{The two original images of the W51-IRS~1 region and their mosaic:
a) Original image of a region 29~arcseconds north of W51~IRS1. b) Original
image of IRS1. c) Mosaic obtained by combining the two previous images.}  
\label{w51c} 
\end{figure} 

\begin{figure}[ht] 
 \begin{center} 
 \begin{minipage}{3.5cm} 
   \begin{tabular}{c} 
\psfig{file=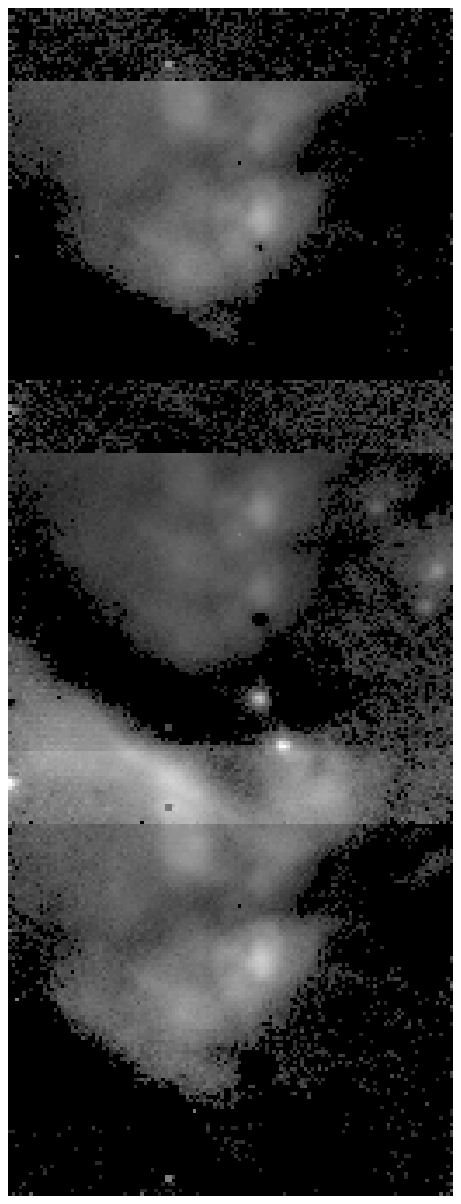,width=3.5cm}\\  a)\\ 
   \end{tabular}  
\end{minipage}
  \begin{minipage}{3.5cm} 
   \begin{tabular}{c} 
\psfig{file=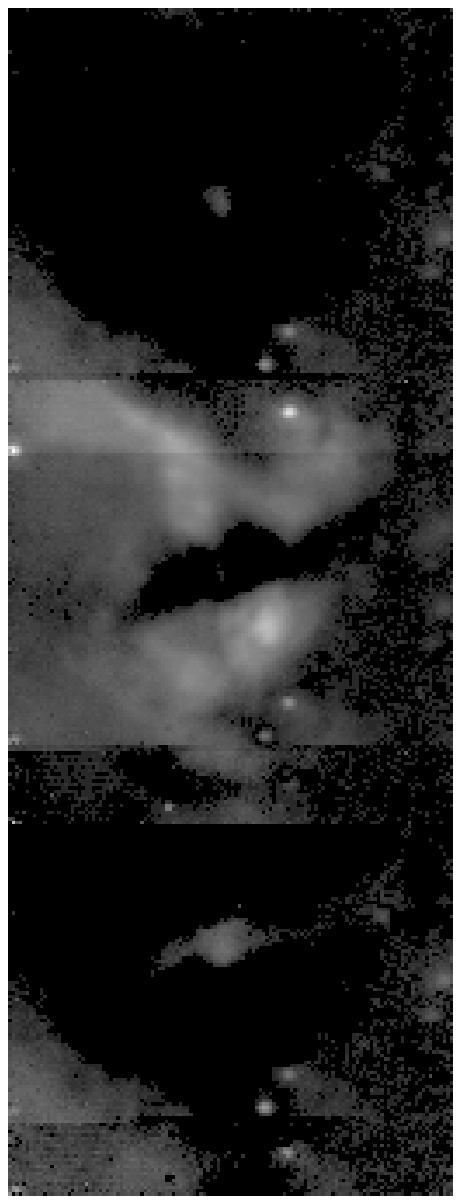,width=3.5cm}\\  b)\\ 
   \end{tabular} 
  \end{minipage}
  \begin{minipage}{3.5cm} 
   \begin{tabular}{c} 
\psfig{file=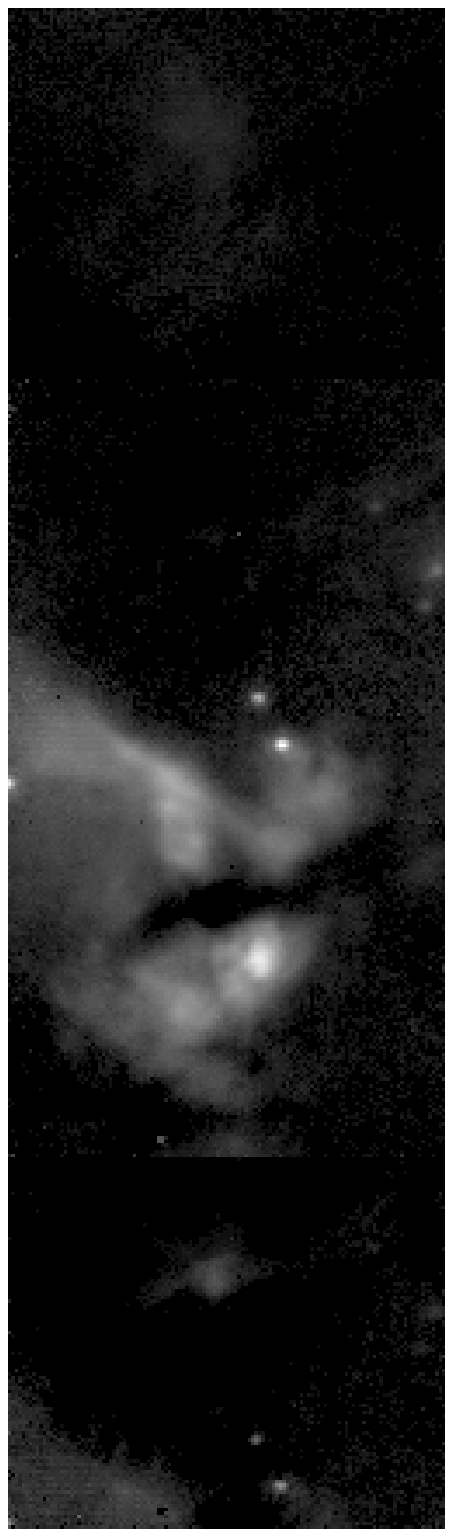,width=3.5cm}\\ c) \\ 
   \end{tabular} 
  \end{minipage}  
 \end{center} 
\caption{Reconstruction of the three images of W51: a) Reconstruction of 
the image of Figure~\protect{\ref{w51c}a)}. b)Reconstruction of 
the image of Figure~\protect{\ref{w51c}b)}. c) Reconstruction of 
the mosaic of Figure~\protect{\ref{w51c}c)} (square root scale 
representation).}  
\label{w51cr} 
\end{figure}

\end{document}